%% file: main.tex
\documentclass[11pt]{iopart}

\usepackage{graphicx}
\usepackage{makecell}

\usepackage{float}
\usepackage[export]{adjustbox}
\usepackage[doi=false,isbn=false, backend=bibtex,style=numeric,natbib=true,sorting=none, maxnames=1]{biblatex}
\usepackage{hyperref}
\usepackage{xcolor}
\hypersetup{
    colorlinks,
    linkcolor={red!50!black},
    citecolor={blue!50!black},
    urlcolor={blue!80!black}
}
 
\usepackage[autostyle=true]{csquotes}
\usepackage{dblfloatfix}
\usepackage[capitalize]{cleveref}  
\usepackage{chemfig}
\usepackage{siunitx}
\usepackage[font=small,labelfont=bf,
   justification=justified,
   format=plain]{caption}

\begin{document}
\title[Tokamak to Stellarator Conversion using Permanent Magnets]{Tokamak to Stellarator Conversion using Permanent Magnets}
\author{M. Madeira$^1$, R. Jorge$^{1,2,3}$}
\address{$^1$ Departamento de Física, Instituto Superior Técnico, Universidade de Lisboa, 1049-001 Lisboa, Portugal}
\address{$^2$ Instituto de Plasmas e Fusão Nuclear, Instituto Superior Técnico, Universidade de Lisboa, 1049-001 Lisboa, Portugal}
\address{$^3$ Department of Physics, University of Wisconsin-Madison, Madison, Wisconsin 53706, USA}
\ead{miguel.madeira@tecnico.ulisboa.pt}
\vspace{10pt}
\begin{indented}
\item[]March 2024
\end{indented}
\begin{abstract}
With the advances in the optimization of magnetic field equilibria, stellarators have become a serious alternative to the tokamak, bringing this concept to the forefront of the pursuit of fusion energy. In order to be successful in experimentally demonstrating the viability of optimized stellarators, we must overcome any potential hurdles in the construction of its electromagnetic coils. Finding cost-effective ways of increasing the number of operating optimized stellarators could be key in cementing this magnetic confinement concept as a contender for a reactor. In this work, an alternative to modular coils, permanent magnets, are studied and are shown to enable the possibility of converting a tokamak into a stellarator. This is then applied to the case of ISTTOK tokamak, where an engineering design study is conducted. 
\end{abstract}

\vspace{2pc}

%\submitto{\jpg}
\maketitle
%\ioptwocol
%%%%%%%%%%%%%%%%%%%%%%%%%%%%%%%%%%%%%%%%%%%%%%%%%%%%%%%%%%%%%%%%%%%%%%
% INTRODUCTION
%%%%%%%%%%%%%%%%%%%%%%%%%%%%%%%%%%%%%%%%%%%%%%%%%%%%%%%%%%%%%%%%%%%%%%
\input{intro}

%%%%%%%%%%%%%%%%%%%%%%%%%%%%%%%%%%%%%%%%%%%%%%%%%%%%%%%%%%%%%%%%%%%%%%
% METHODOLOGY
%%%%%%%%%%%%%%%%%%%%%%%%%%%%%%%%%%%%%%%%%%%%%%%%%%%%%%%%%%%%%%%%%%%%%%

\input{methodology}

%%%%%%%%%%%%%%%%%%%%%%%%%%%%%%%%%%%%%%%%%%%%%%%%%%%%%%%%%%%%%%%%%%%%%%
% RESULTS
%%%%%%%%%%%%%%%%%%%%%%%%%%%%%%%%%%%%%%%%%%%%%%%%%%%%%%%%%%%%%%%%%%%%%%

\input{results}

%%%%%%%%%%%%%%%%%%%%%%%%%%%%%%%%%%%%%%%%%%%%%%%%%%%%%%%%%%%%%%%%%%%%%%
% CONCLUSIONS
%%%%%%%%%%%%%%%%%%%%%%%%%%%%%%%%%%%%%%%%%%%%%%%%%%%%%%%%%%%%%%%%%%%%%%

\input{conclusions}

%%%%%%%%%%%%%%%%%%%%%%%%%%%%%%%%%%%%%%%%%%%%%%%%%%%%%%%%%%%%%%%%%%%%%%
% ACKNOWLEDGMENTS
%%%%%%%%%%%%%%%%%%%%%%%%%%%%%%%%%%%%%%%%%%%%%%%%%%%%%%%%%%%%%%%%%%%%%%

\input{acknowledgements}

%%%%%%%%%%%%%%%%%%%%%%%%%%%%%%%%%%%%%%%%%%%%%%%%%%%%%%%%%%%%%%%%%%%%%%
% REFERENCES
%%%%%%%%%%%%%%%%%%%%%%%%%%%%%%%%%%%%%%%%%%%%%%%%%%%%%%%%%%%%%%%%%%%%%%

%\renewcommand*{\bibfont}{\scriptsize}
%\printbibliography[heading=secbib]
\printbibliography
%%%%%%%%%%%%%%%%%%%%%%%%%%%%%%%%%%%%%%%%%%%%%%%%%%%%%%%%%%%%%%%%%%%%%%
\end{document}

%% file: intro.tex
\section{Introduction}
\label{chapter:introduction}

Successive breakthroughs in stellarator research have greatly reduced what was considered to be their primary drawback, neoclassical transport \cite{Helander2014}. This has been experimentally realized with the W7-X device \cite{Beidler2021}. Furthermore, due to having no plasma current, stellarators offer inherent steady-state operation and far fewer magnetohydrodynamic (MHD) instabilities, leading to no disruptions. On the other hand, they are technically more complex, usually relying on non-planar coils to generate their magnetic field. 

The optimization, construction, and assembly of complex coil systems have proven to be one of the biggest bottlenecks and cost drivers in the development of modern stellarators \cite{Bosch2013,Lobsien2018, Lion2021}. For instance, in the case of W7-X, coil manufacturing tolerances were approximately $0.1\%$ relative to the coil radius, while the magnetic system placement tolerances were kept below $\SI{1.5}{\centi\meter}$ \cite{Bosch2013}. The
maximum coil assembly error remained under \SI{4.4}{\milli\meter} \cite{Bosch2013}. Meeting these stringent requirements
was only possible with precision metrology and continual adjustments of the built components, including a re-optimization procedure based on the up-to-date coil and module shapes and positions \cite{Bosch2013}.
This has sparked a search for simpler methods of achieving the desired magnetic field configurations without compromising field quality and plasma confinement. 

Although magnetic confinement devices have traditionally relied on electromagnets to create a toroidal magnetic field equilibrium, Ref. \cite{Helander2020} has introduced the idea of coupling coils with permanent magnets (PMs). By providing additional three-dimensional shaping of the magnetic field, PMs can alleviate the role of non-planar coils, thus allowing a reduction in the complexity of the coil systems Refs. \cite{Helander2020, Zhu2020, Zhu2020topology, Zhu2022, Qian2022, Xu2021}. For less stringent magnetic equilibria, PMs are even able to eliminate the need for non-planar coils allowing for a stellarator design that requires only toroidal field coils similar to those of a tokamak \cite{Zhu2020, Zhu2020topology, Zhu2022, Qian2022, Xu2021, Lu2022, Yu2024}.

Besides simplifying the coil systems, permanent magnets offer other advantages. PMs have a steady magnetic field that requires no external source of power, they only require passive mechanical support and there is no need for cryogenic electrical connections, in contrast with superconducting coils \cite{Coey, Coey_Parkin_2021, Jiles_2017}. Furthermore, the materials used in permanent magnets are inexpensive, especially when compared to the superconductors used in the modular coils of fusion devices. The cost of assembly and the supporting structure likely exceeds the material cost of the magnets \cite{Zhu2020}. 
This makes PMs a cost-effective choice, reducing both manufacturing and energy consumption expenses. 
Finally, PMs would be able to produce less field ripple than modular coils due to having a more continuous distribution \cite{Kaptanoglu2022}. 
However, permanent magnets have some disadvantages: their inability to be turned off, their inherently discrete and limited magnetic field strength, and the prospect of demagnetization \cite{Coey}. 

Fortunately, magnetic materials have considerably improved, with the magnetic field energy density doubling every 12 years throughout the 20th century \cite{Coey}. Commercial magnets can now have remnant magnetic fields $B_r$ larger than $\SI{1.4}{\tesla}$ while also having a coercive field $\mu_0H_c$ larger than $\SI{2.2}{\tesla}$ at room temperature \cite{Li2018}. Coercivity may be increased to as high as $\SI{3.5}{\tesla}$ albeit with a reduction in the remnant field \cite{Li2018}. Coercivity may be further increased by reducing the magnet's temperature. A sintered neodymium magnet with a coercive field of $\mu_0H_c = \SI{1.25}{\tesla}$ at room temperature has a much higher $\mu_0H_c > \SI{5}{\tesla}$ at \SI{77}{\kelvin} \cite{Durst1987}. In a similar manner to how high-temperature superconductor coils are cooled down to liquid nitrogen temperature (\SI{77}{\kelvin}) to enable the generation of very large magnetic fields, magnets could be cooled down to prevent their demagnetization.
These properties make PMs relevant for most of 
 the present-day magnetic confinement facilities, especially for university-scale projects.

Permanent magnets are present in a large number of high-field applications, namely particle accelerators \cite{Thonet2016}, electric motors \cite{Zhu2001}, and magnetic resonance imaging \cite{Blumich2008}. Nonetheless, for a fusion reactor, where the average toroidal field is likely to exceed \SI{5}{\tesla} \cite{Lion2021}, magnetic materials need to either be improved or restricted to the low field side. Alternatives that still produce dipole-like fields without relying on magnetic materials, such as superconducting tiles \cite{Neilson2011} or small superconducting saddle coils could represent a way of extending the core principles of PM stellarators to higher field machines.

After the demonstration of the possibility of designing stellarators with permanent magnets by Ref. \cite{Helander2020} and the realization that present-day materials are relevant for fusion research, large efforts began in adapting existing codes \cite{Zhu2020} and developing new codes for PM design, namely FAMUS \cite{Zhu2020topology}, MAGPIE \cite{Hammond2020} and SIMSOPT \cite{Kaptanoglu2022, SIMSOPT, Kaptanoglu2022relax}. Additionally, the PM4Stell project \cite{Zhu2022} attempted to prove the technical viability of permanent magnets through the construction of a half-period device that uses the NCSX equilibrium, replacing the modular coils with a permanent magnet structure. Finally, a small quasi-axisymmetric stellarator, MUSE, has been built using only normally oriented PMs and circular toroidal field coils \cite{Qian2022}. The present work leverages the outlined projects and makes use of their developed tools, especially the greedy optimization algorithms from SIMSOPT \cite{Kaptanoglu2022} and MAGPIE \cite{Hammond2020} (see \cref{sec:meth}).

Historically, tokamaks have yielded larger energy confinement times than stellarators, while being technically simpler. This has led to tokamaks outnumbering stellarators roughly by a factor of 5 \cite{IAEA2022}. Increasing the experimental efforts, even with smaller machines, could greatly increase the understanding of different stellarator equilibria and plasma regimes. Nonetheless, this requires a large investment or cheaper alternatives for designing and building stellarators.
Conversely to how the model C stellarator was repurposed to be the Symmetric Tokamak \cite{stix1998highlights}, in this work, we propose the conversion of a tokamak into a stellarator.
This idea arises from the realization that by swapping the transformer/solenoid of a tokamak with permanent magnets, we have all the necessary components to build a stellarator without a large investment.
Although a large portion of the obtained results should be general and applicable to other tokamaks, for concreteness, we applied the tools developed in this work to the conversion of the ISTTOK tokamak \cite{Varandas1996} to a stellarator.

This paper is organized as follows. \cref{chapter:ISTTOK} provides an overview of the ISTTOK tokamak, its current setup, and the aspects to take into account for adding permanent magnets. In \cref{sec:meth}, the used permanent magnet optimization and array design methods are described. Core design choices, namely the omnigeneous type of magnetic equilibrium (quasi-axisymmetric, QA; quasi-helically symmetric, QH; or quasi-isodynamic, QI), the vacuum vessel shape, the magnet array design, and the magnet's orientations are addressed in \cref{chapter:designchoices}. In \cref{section:ISTELL}, we optimize a suitable equilibrium and evaluate three different scenarios to identify a viable design for converting the ISTTOK tokamak into a stellarator. The conclusions follow.

\section{Description of the ISTTOK tokamak}
\label{chapter:ISTTOK}

ISTTOK is the only toroidal magnetic confinement device in Portugal. Its construction began in 1990 with many of its components being repurposed from the TORTUR tokamak \cite{Varandas1996}.
Its toroidal magnetic field results from a set of 24 circular coils which produce an on-axis magnetic field of $\SI{0.5}{\tesla}$. These coils could generate up to $\SI{3}{\tesla}$ if coupled to a cooling system and a larger power
source. The poloidal magnetic field is provided by four vertical and two horizontal field coils beyond the self-created field by the plasma current. Today, its main objectives are the study of plasma turbulence, operation and control on alternating plasma current regimes, testing liquid metal limiter concepts, developing and upgrading plasma-relevant diagnostics for nuclear fusion, and supporting nuclear fusion and plasma physics education.

\begin{figure}[h]
  \begin{minipage}[b]{.4\linewidth}
    \centering
    \captionof{table}{ISTTOK experimental parameters.}
    \small
    \begin{tabular}{@{}ll}\br
    VV Major Radius ($R_0$) &\SI{46}{\centi\meter} \\
    VV Minor Radius ($a$) &\SI{8.5}{\centi\meter} \\
    Copper Shell Thickness &\SI{1.5}{\centi\meter} \\
    Magnetic Field &\SI{0.5}{\tesla} \\
    TF Coil Radius &\SI{20.25}{\centi\meter} \\
    TF Coil Axis Radius &\SI{52}{\centi\meter} \\
    Plasma Current &\SI{7}{\kilo\ampere} \\
    Electron Temperature &\SI{120}{\electronvolt} \\
    Plasma Density &\SI{5e18}{\per\meter\cubed} \\\br
    \end{tabular}
    \label{tab:ISTTOK}
  \end{minipage}\hfill \hspace{5mm}
  \begin{minipage}[b]{.6\linewidth}
    \centering
    \includegraphics[width = .85\linewidth, valign = c]{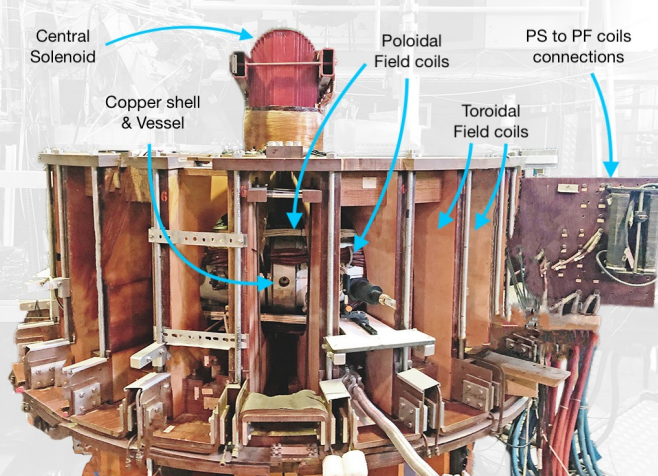}
    \captionof{figure}{ISTTOK's main components \cite{CoronaRivera}.}
    \label{fig:ISTTOK}
  \end{minipage}
\end{figure}

The experimental parameters of ISTTOK relevant for this work are summarised in \cref{tab:ISTTOK} and \cref{fig:ISTTOK} showcases its main components. With a magnetic field of \SI{0.5}{\tesla} at a major radius of \SI{0.46}{\meter} and a coil radius of \SI{0.2025}{\meter}, the maximum field inside the nominal toroidal field coils is $\leq \SI{0.9}{\tesla}$ which is comfortably below the coercive field of grade N52 \chemfig{NbFeB} magnets. Furthermore, the PM4Stell project tested demagnetization effects also using a $\SI{0.5}{\tesla}$ toroidal field for NCSX, whose coil radius is considerably larger, and calculated that less than $1\%$ of the magnets were at risk of demagnetization \cite{Zhu2022}.  

\vspace{-5mm}

\begin{figure}
    \centering
    \includegraphics[width =.5\linewidth]{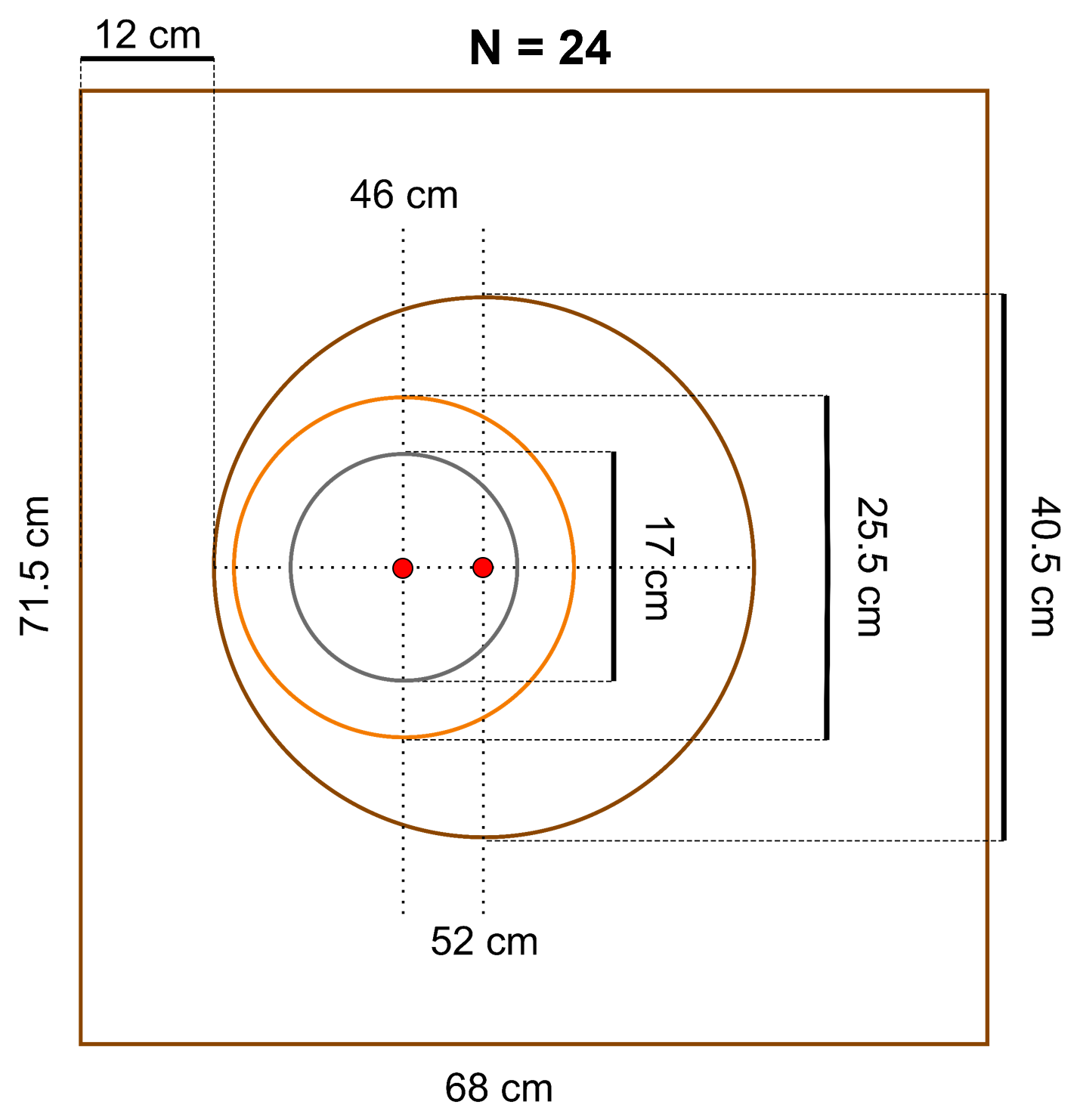}
    \caption{Cross section of ISTTOK's current setup. There are $N = 24$ toroidal field coils (brown). The vacuum vessel has a minor radius of $\SI{8.5}{\centi\meter}$ in its narrowest section (grey). The vessel is enveloped by a copper shell that maintains a minor radius of $\SI{10}{\centi\meter}$ for the majority of its extent. However, its widest section, corresponding to the region where the two halves of the vessel are fixed together, has a radius of $\SI{12.75}{\centi\meter}$ (orange).}
    \label{fig:ISTTOKshifted}
\end{figure}

A cross-section of ISTTOK's current setup is illustrated in \cref{fig:ISTTOKshifted}. The concept of transforming ISTTOK into a stellarator using permanent magnets stems from the recognition that there is a substantial volume available between the coils and the copper shell enveloping the vacuum vessel. The copper shell, for the majority of its extent, maintains a minor radius of $\SI{10}{\centi\meter}$, while the coils have a circular shape with a radius of $\SI{20.25}{\centi\meter}$. This configuration allows for the placement of magnets close to the plasma, thereby maximizing their impact. If the coils were directly positioned on the vessel, it would only be feasible to place magnets at a relevant distance to the plasma between the coils. Nonetheless, this configuration still poses challenges, particularly since the available volume on the high field side is significantly smaller than on the low field side. It is anticipated that a larger volume of magnets will be required where the toroidal field is the largest. This challenge is further explored in \cref{section:ISTELLscenarios}.

\begin{figure}
    \centering
    \includegraphics[width =.7\linewidth]{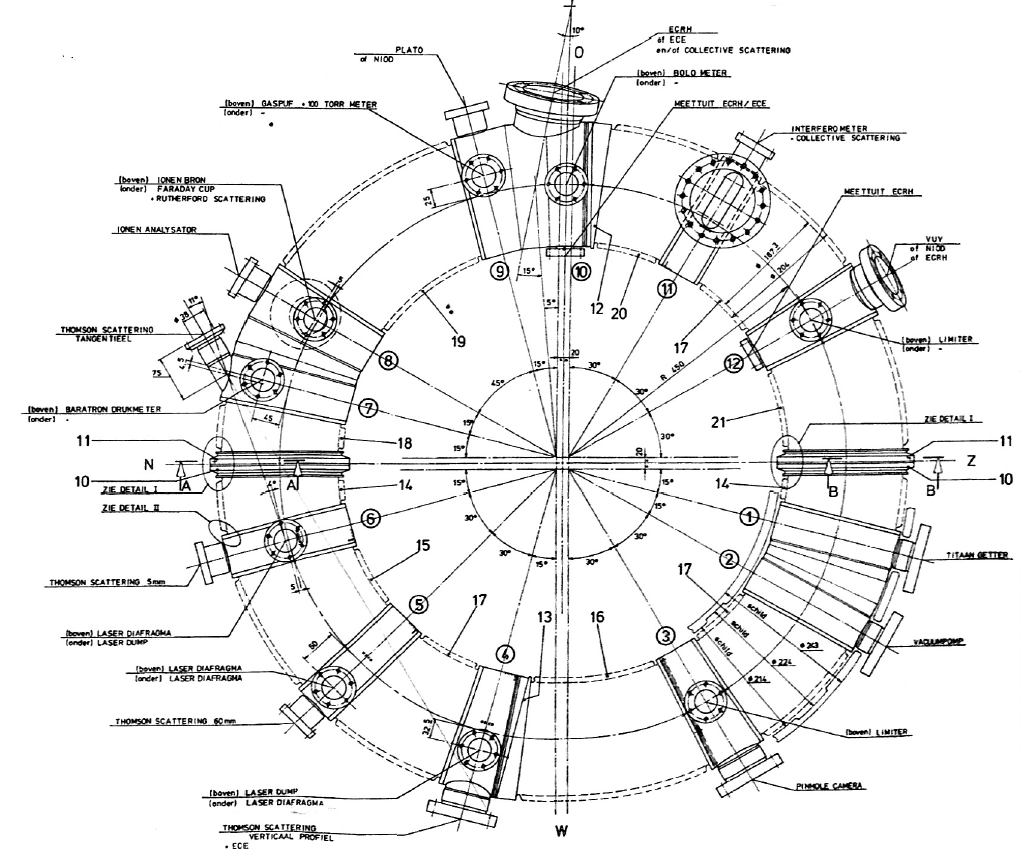}
    \caption{Schematic of the vacuum vessel and plasma accessing ports of ISTTOK. Note that all ports are to good approximation cylindrical in shape. The ports adhere to one of three flange standards, DN40, DN60, or DN100. The functions of each port are merely illustrative and do not correspond to the current use of the ports.}
    \label{fig:ISTTOKscheme}
    \vspace{-4mm}
\end{figure}

Another aspect to take into account in ISTTOK's current design is the plasma accessing ports, see \cref{fig:ISTTOKscheme}. The plasma accessing ports pose a challenge to the integration of permanent magnets in magnetic confinement devices, directly influencing the available volume for magnet placement. In the design of a machine from scratch, the port placement can be strategically chosen after determining ideal magnet locations, and the number of ports can be minimized. Furthermore, the locations of these ports can be selected to adhere to stellarator and field period symmetry, see Ref. \cite{Qian2023} and \cref{section:diffVV}. For ISTTOK however, if the vacuum vessel is to be kept, the large number of ports, 33 in total, the lack of either of the mentioned symmetries, and the need for the magnet placement to accommodate the existing port locations present an added layer of complexity. One positive characteristic of ISTTOK's ports is their geometrical simplicity. All ports are cylindrical and their dimensions fall on one of three flange standards, DN40, DN60, or DN100. The modeling and impact of including the ports in the design are evaluated in \cref{section:ports}.

%% file: methodology.tex
\section{Optimization Methods}
\label{sec:meth}

In this work, we use the SIMSOPT code \cite{SIMSOPT} for magnetic equilibrium optimization and permanent magnet optimization.
The goal of permanent magnet optimization is to find a magnetization distribution that produces an intended magnetic field. This field is usually computed in one of two ways, effective currents or discretized magnetic dipoles. We use the latter approach. The available volume for the magnetic material is split into elements of a mesh, with the center of each element depicting a dipole. By optimizing a predefined set of possible available positions, the complexity of the problem is considerably reduced, while ensuring the solution obtained can be reasonably built.

Two types of magnet arrays were used, a uniform toroidal grid, following the scheme in Ref. \cite{Kaptanoglu2022} and the trapezoidally-enclosed rectangular prisms (\textit{trec}) \cite{Hammond2022} design from MAGPIE \cite{Hammond2022}.
For the first, the array is defined between two toroidally closed surfaces. The radial grid spacing, $dr$, is an input, while both the toroidal and poloidal grid units are defined by the quadrature points of the plasma surface. In general, a $64 \times 64$ quadrature grid was considered per half-field period. 
For the second, each magnet is a rectangular prism that conforms to a rectangular grid in cylindrical coordinates. The magnet array consists of drawer-like bundles that are grouped into a number of toroidal wedges. Within a drawer, the magnets are glued to one another. Each wedge is divided into two subsets, one for the high field side and one for the low field side, which is further determined separately at the top and bottom to best adjust the inner limiting surface. In this concept, all magnets have the same shape and size, there are no unique magnets, which reduces magnet and assembly cost, while facilitating the array construction (see \cref{fig:RealisticGrid} and Refs. \cite{Zhu2022, Hammond2022}).

Having defined the magnet mesh, similarly to coil optimization, our objective is to minimize the squared error field on the plasma surface.
The optimization problem can then be formulated as
\begin{eqnarray}
  {\min}   &&    f_B = \int_{\partial V} \left[\left(\textbf{B}_M (\textbf{m}) + \textbf{B}_\text{coils} \right)\cdot \textbf{n} \right]^2 d\partial V,            
  \nonumber           \\
  {\rm w.r.t. }     && {\bf m},\\                              
\text{where }   && \textbf{B}_M = \frac{\mu_0}{4\pi} \sum_{j=1}^D\left(3\frac{(\textbf{m}_j \cdot \textbf{r}_j)}{r_j^5}\textbf{r}_j  - \frac{\textbf{m}_j}{r_j^3}\right),
  \nonumber 
  \label{eq:optproblem}
\end{eqnarray}
with $\textbf{n}$ the normal vector to the magnetic surface, $\textbf{m}$ a vector with the magnetic moment of all the $D$ dipoles, and $\textbf{B}_\text{coils}$ is the field produced by the coil system. The objective function $f_B$ can be reformulated as \cite{Kaptanoglu2022relax}
\begin{equation}
    f_B = ||\mathbf{A}\mathbf{m}-\mathbf{b}||_2^2 ,
    \label{eq:objfunction}
\end{equation}
where $\mathbf{A}$ is a matrix that depends only on the geometry of the problem and $\mathbf{b}$ encapsulates the magnetic field from sources other than the permanent magnets. These two factors can be computed before the optimization begins, and then re-used, thus making this form of $f_B$ more computationally efficient. The magnetic material used in the design will be evaluated by the effective volume,  
\begin{equation}
    V_{eff} = \sum_{j=1}^D \frac{||m_j||_2}{M}.
\end{equation}
One engineering restriction that is essential in permanent magnet optimization is a limit on the value of the magnetization,
\begin{equation}
    ||\mathbf{m}||^2 \leq (m_i^{\max})^2 .
    \label{eq:maxM}
\end{equation}
This guarantees that the calculated magnet solution can be constructed from real magnets. In the present work, we consider commercial N52 neodymium-iron-boron magnets \cite{Risheng} for their high remnant fields, $B_r = 1.43-1.48\SI{}{\tesla}$ ($m_i^{\max} = B_rV_i/\mu_0$, where $V_i$ is the volume of magnet $i$), and their high coercive fields, $H_c \leq \SI{2}{\tesla}$.

Having defined \cref{eq:objfunction} with the restriction in \cref{eq:maxM}, the most common approach is to perform continuous optimizations that take advantage of gradient-based algorithms \cite{Helander2020, Zhu2020topology,Kaptanoglu2022relax,Qian2022,Hammond2022}. Choosing this approach has the consequence of producing solutions where each cell has a different magnetic moment with possibly varying orientations. Different magnetic moments can be attainable by either using distinct materials, effectively changing the magnetization or by using magnets with different volumes. This would make manufacturing and assembly of magnet grids highly non-efficient, defeating the purpose of introducing PMs in stellarators. Therefore, the solutions need to be further tuned to produce binary dipole distributions with a fixed magnetization and the magnet orientations need to be mapped to a set of possibilities to ensure they are feasible. Those two steps reduce the field accuracy of the found solution.

In Ref. \cite{Kaptanoglu2022}, the \textbf{G}reedy \textbf{P}ermanent \textbf{M}agnet \textbf{O}ptimization (\textbf{GPMO}) algorithm that directly produces binary dipole arrays from a set of possible orientations was introduced. Using this algorithm we can define the material to be used, the set of possible magnet orientations, and the magnet grid, before the optimization. This is the ideal case, as it keeps the number of unique magnets to a minimum, facilitating the fabrication and assembly of the magnet system. 
Furthermore, GPMO outperforms other state-of-the-art algorithms while being substantially faster \cite{Kaptanoglu2022}. Due to these factors, GPMO was chosen as the method to use in this work. 
Since the solutions are binary, the dipole moment is normalized by its respective maximum in the previous formulation (see Ref. \cite{Kaptanoglu2022}).

The GPMO algorithm can be outlined in the following way. The maximum-strength magnetic moment that minimizes $f_B$ out of a predefined mesh of possible dipole positions and orientations is added individually at each iteration. Once a magnet is placed, its corresponding position is removed from the grid of available positions. The procedure repeats until $K$ magnets have been positioned.  
We note that the GPMO baseline algorithm has been expanded to include other variants, such as the baseline GPMO, GPMO backtracking (GPMOb) and its \textbf{Arb}itrary \textbf{Vec}tor (ArbVec) versions \cite{Kaptanoglu2022, Hammond2023}.
GPMOb works by backtracking every $K_b$ iteration, checking the $N_{\text{adjacent}}$ adjacent magnets for each magnet placed and removing the oppositely orientated pairs it finds. GPMOb is more accurate than the baseline version and requires fewer magnets, thus GPMOb solutions are more cost-effective. The ArbVec version of the baseline and backtracking algorithms introduces user-specified magnet orientations. Before this addition, the magnet orientations were defined by the coordinate system. Therefore, when using a uniform toroidal grid, the magnets followed the toroidal, poloidal, or radial directions. For further descriptions of GPMO see \cite{Kaptanoglu2022} and \cite{Hammond2023}.

%% file: results.tex
\section{Design Choices: Equilibrium, Vessel, and Magnet Grid}
\label{chapter:designchoices}

Before establishing a definite solution, it is enlightening to start with a simpler approach, namely to use uniform grids to assess the viability of the project and to evaluate multiple early design choices. 

\subsection{Magnetic Configuration}
\label{section:uniformgridISTELL}

First, two aspects were considered - the type of magnetic field omnigenity (QA, QH, or QI) and the impact of the toroidal magnetic field intensity on the resulting field quality. When optimizing electromagnetic coils only, the latter is typically not a concern, due to the fact that to reach higher fields, one simply needs to scale the current up. In contrast, for permanent magnets, the field intensity may only be altered by either changing the magnets' volume and/or its material. Ideally, when designing a magnet grid, all magnets have a predetermined volume, are restricted to one material only, and their orientations are chosen from a finite set. This makes construction and assembly considerably easier by keeping the number of unique magnets to a minimum. As the average toroidal magnetic field increases, to keep a given magnetic equilibrium the poloidal component must increase accordingly. However, given the fact that the maximum volume of the magnet mesh is fixed, the field produced by the magnets will, at some point, be too low to keep up with the toroidal field intensity produced by the coils.

To confirm this idea, three different configurations were used: a precise QA configuration \cite{miguel_madeira_2024_10656980} with  $\iota_{QA} = 0.185$ and a number of field periods of $\mathrm{nfp}_{QA} = 2$, the precise QH from Ref. \cite{Landreman2021} with $\iota_{QH} = 1.24$ and $\mathrm{nfp}_{QH} = 4$, and W7-X's standard configuration \cite{Geiger2015} with $\iota_{W7-X} = 1$ and $\mathrm{nfp}_{W7-X} = 5$ \cite{Bosch2013}. The major radius was chosen to match ISTTOK's, $R_0 = \SI{0.46}{\meter}$. 

Permanent magnet optimizations using the baseline GPMO algorithm were performed scanning over different values of toroidal magnetic flux, $\phi_{edge}$, for the last closed flux surface,\vspace{-1mm}
\begin{equation}
    \phi_{edge} = \int_S \textbf{B}\cdot\textbf{n }dS,
\end{equation} 
where $S$ is the surface delimited by the plasma boundary. The magnet grids were defined in toroidal coordinates with 64 by 64 quadrature points and a magnet radial dimension of $dr = \SI{2}{\centi\meter}$ following the SIMSOPT scheme explained in \cref{sec:meth}. The grid's inner and outer limiting surfaces were separated by a radial extent of $r_{ext} = \SI{13}{\centi\meter}$. The toroidal component of the magnetic field is provided by 24 circular coils whose currents are optimized before the PM optimization.

\begin{figure}
    \centering 
    \includegraphics[width = 0.6\textwidth, valign=c]{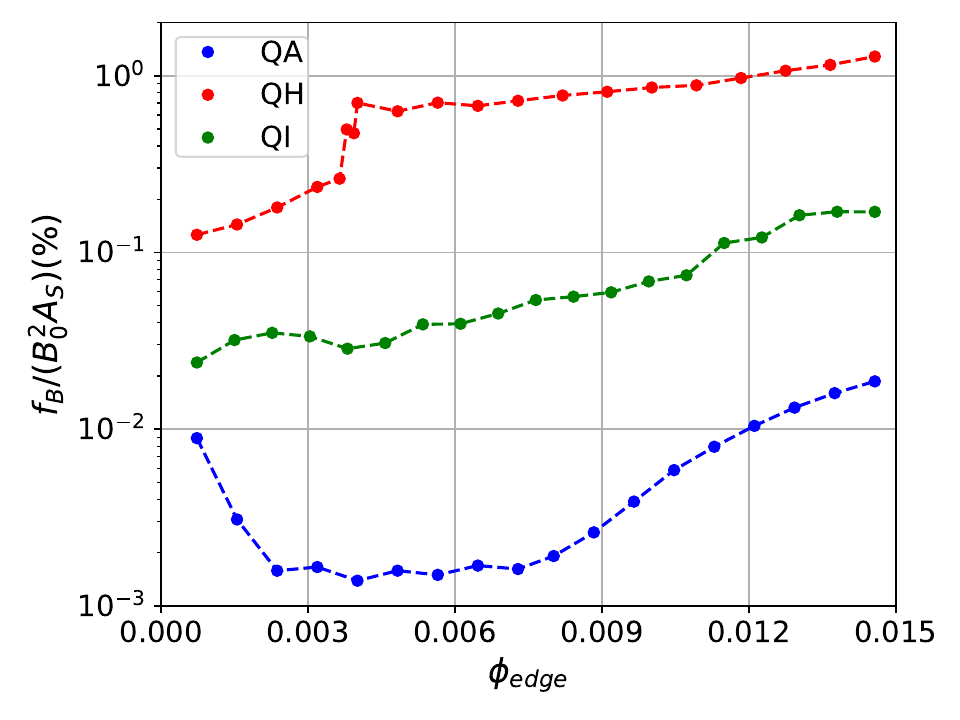}
    \captionsetup{width=\textwidth}
    \vspace{-5mm}
    \caption{Squared normal field objective function normalized by the square of the field intensity at the axis, $B_0^2$, and the surface area of the equilibrium, $A_S = 4\pi R_{0}^2/(2A\mathrm{nfp})$, as a function of $\phi_{edge}$ for three magnetic field surfaces. The equilibria are more easily reproduced in the order QA, QI, and QH, with the QA configuration being the only one with an error field below $0.1\%$.}
    \label{fig:ScaleStudydata}
\end{figure}

The obtained objective function, $f_B$, for a given $\phi_{edge}$ value is represented in \cref{fig:ScaleStudydata}.
In order to properly assess if the field quality is relatively increasing as $\phi_{edge}$ decreases, the objective function is normalized by the square of the field intensity at the axis, $(B_0)^2$ multiplied by the surface area of the equilibrium, $A_S = 4\pi R_{0}^2/A$, where $A = R_0/a$ is the aspect ratio, $R_0$ is the major radius and $a$ is the minor radius. 

As shown in \cref{fig:ScaleStudydata}, the equilibria are more easily reproduced in the order: QA, QI, QH. This aligns with the increasing order of $\iota$. Notably, low $\iota$ QA configurations prove to be the most achievable with permanent magnets, with the QA configuration being the only one with an error field below $0.01\%$. This is expected as higher $\iota$ requires a higher poloidal field strength, which is provided by the PMs. Additionally, QA configurations are more convenient for fitting into a circular vessel, making it the selected magnetic equilibrium used in this work.

Although the behavior of the curves for the three configurations is vastly different, a noticeable trend emerges at high $\phi_{edge}$. Namely, the relative error field increases with increasing magnetic field flux. However, for the QA case, between $\phi_{edge} \approx \SI{0.002}{\tesla\per\meter\squared}$ and $\phi_{edge} \approx \SI{0.008}{\tesla\per\meter\squared}$, there is a plateau. This suggests that the magnetization of the considered material (N52 \chemfig{NdFeB}) is enough to obtain the poloidal component of the field and no relative advantage is gained from reducing the toroidal magnetic field. For very low magnetic field flux, there is an increase that may suggest the magnetization is too high, and using weaker magnets could be beneficial. 

\subsection{Vacuum Vessel Shape}
\label{section:VVshape}

We now assess the influence of the shape of the vacuum vessel (VV) on the optimization of the permanent magnets. In tokamaks, a circular or D-shaped VV is commonly employed, while for stellarators, the VV usually follows the cross-section of the nested magnetic surfaces. We then considered two grids, one with a circular cross-section and one following the cross-section of the plasma boundary.

The grids followed a setup similar to the one used above, namely toroidal coordinates, 64x64 quadrature points, $dr = \SI{2}{\centi\meter}$, $d_{plasma} = \SI{2}{\centi\meter}$ and 24 circular coils. The toroidal magnetic field was kept at $\SI{1}{\tesla}$. For the circular grid, a radial extent of $r_{ext} = \SI{13}{\centi\meter}$ was defined, while for the magnetic surface cross-section, two cases were investigated, one with the same radial extent as the circular case and one where $r_{ext}$ was chosen to match the effective permanent magnet volume, $V_{eff}$, of the circular case. This bifurcation arose from the recognition that for the same radial extent, there was a larger effective volume available to the circular case. 

\begin{figure}
    \centering 
    \includegraphics[width =0.55\textwidth,valign=c]{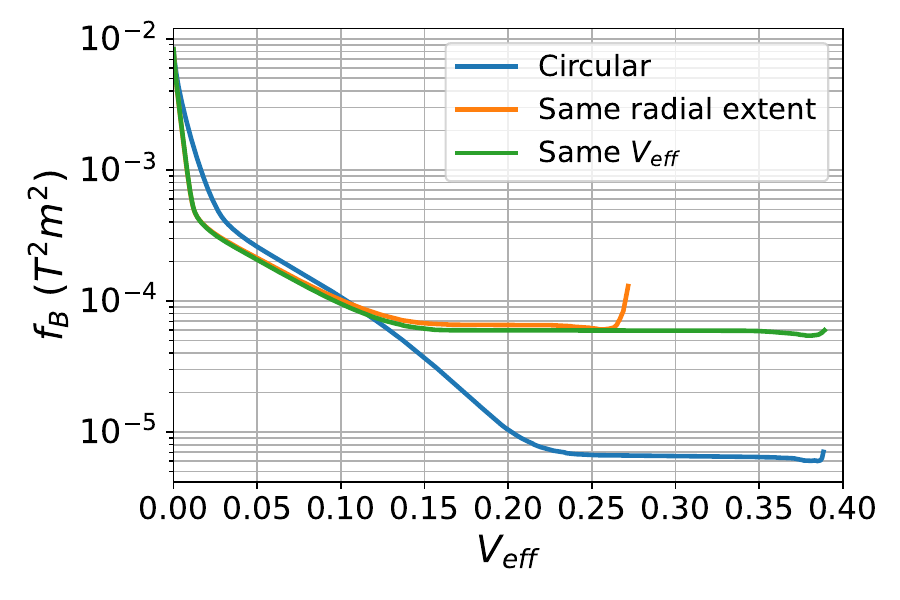}
    \includegraphics[width =0.44\textwidth,valign=c]{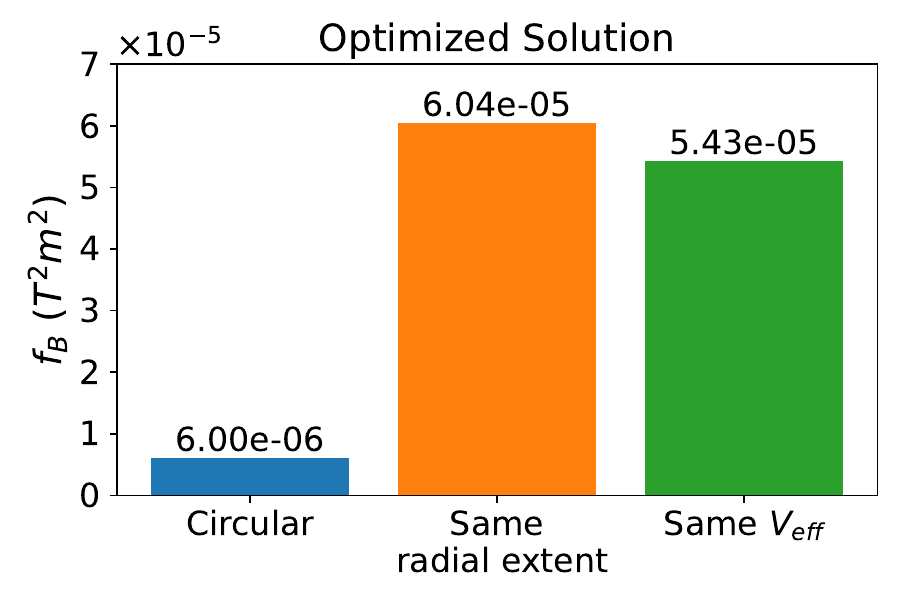}
    \captionsetup{width=\textwidth}
    \vspace{-5mm}
    \caption[Optimization curves and minima for a QA configuration using circular and plasma-shaped cross sections.]{Left: Optimization curves for a QA configuration using a circular cross-section VV and the equilibrium's cross-section VV with the same radial extent or maximum effective volume as the circular case. The circular cross-section leads to a more optimized solution, with a lower cost function, although it requires a larger effective volume to reach a plateau. Right: Bar chart of the solution with the minimum cost function for each case.}
    \label{fig:uniform_grid}
\end{figure}

\begin{figure}
    \centering 
    \includegraphics[width =\textwidth]{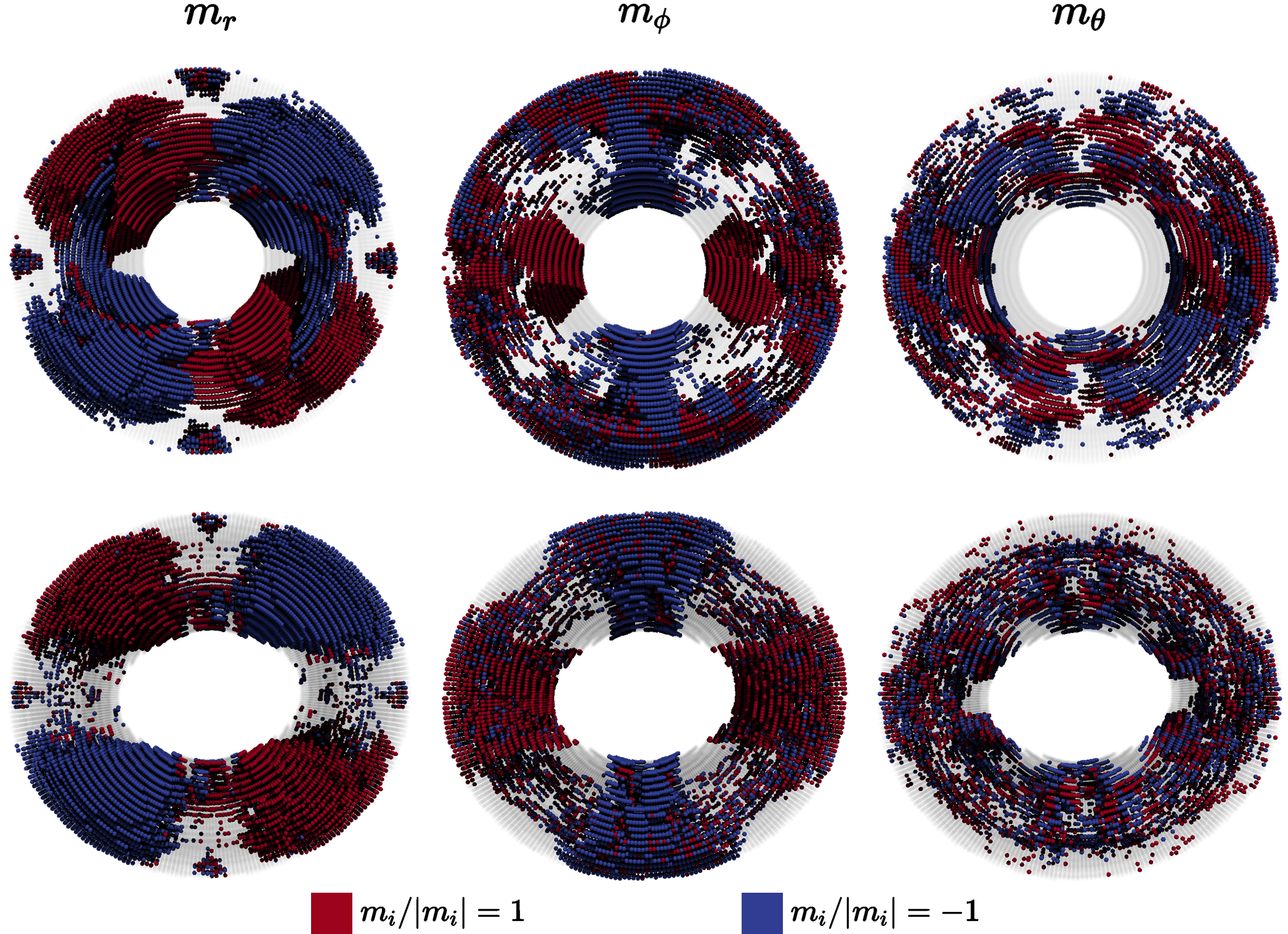}
    \captionsetup{width=\textwidth}
    \caption[magnet configurations in the radial, toroidal and poloidal directions for uniform grids with circular and plasma shaped cross sections.]{Top view of the magnet configurations in the radial, toroidal and poloidal directions for uniform grids with circular cross-section (top) and with an equilibrium shaped cross section with the same $r_{ext}$ (bottom) and with the same $V_{eff}$ (bottom). There are 21328 and 14190 magnets respectively, corresponding to the optimal solution obtained by the baseline GPMO algorithm with toroidal orientations. The colors represent the normalized magnetization direction. 
    }
    \label{fig:UniformGridResults}
    \vspace{-4mm}
\end{figure}

The three described cases were optimized for the same QA configuration using the baseline GPMO algorithm. The resulting optimization curves are shown in \cref{fig:uniform_grid} and the magnet configurations for the optimized solutions are shown in \cref{fig:UniformGridResults}.
From the optimization curves in \cref{fig:uniform_grid} left, it is clear that using a circular VV is advantageous when compared to using a VV with the equilibrium's magnetic surface cross-section, as $f_B$ is roughly one order of magnitude lower. Although for the latter, the cost function initially decreases faster with the used magnet volume, it also reaches a plateau first, around $V_{eff} \sim \SI{0.15}{\meter\cubed}$ compared to $V_{eff} \sim \SI{0.25}{\meter\cubed}$. This is a positive result for converting tokamaks into stellarators, as it means that, at least for low $\iota$ QA configurations, the VV may be kept the same without a loss in field quality, as long as multiple magnet orientations are considered.

\begin{figure}
    \centering 
    \includegraphics[width =\textwidth]{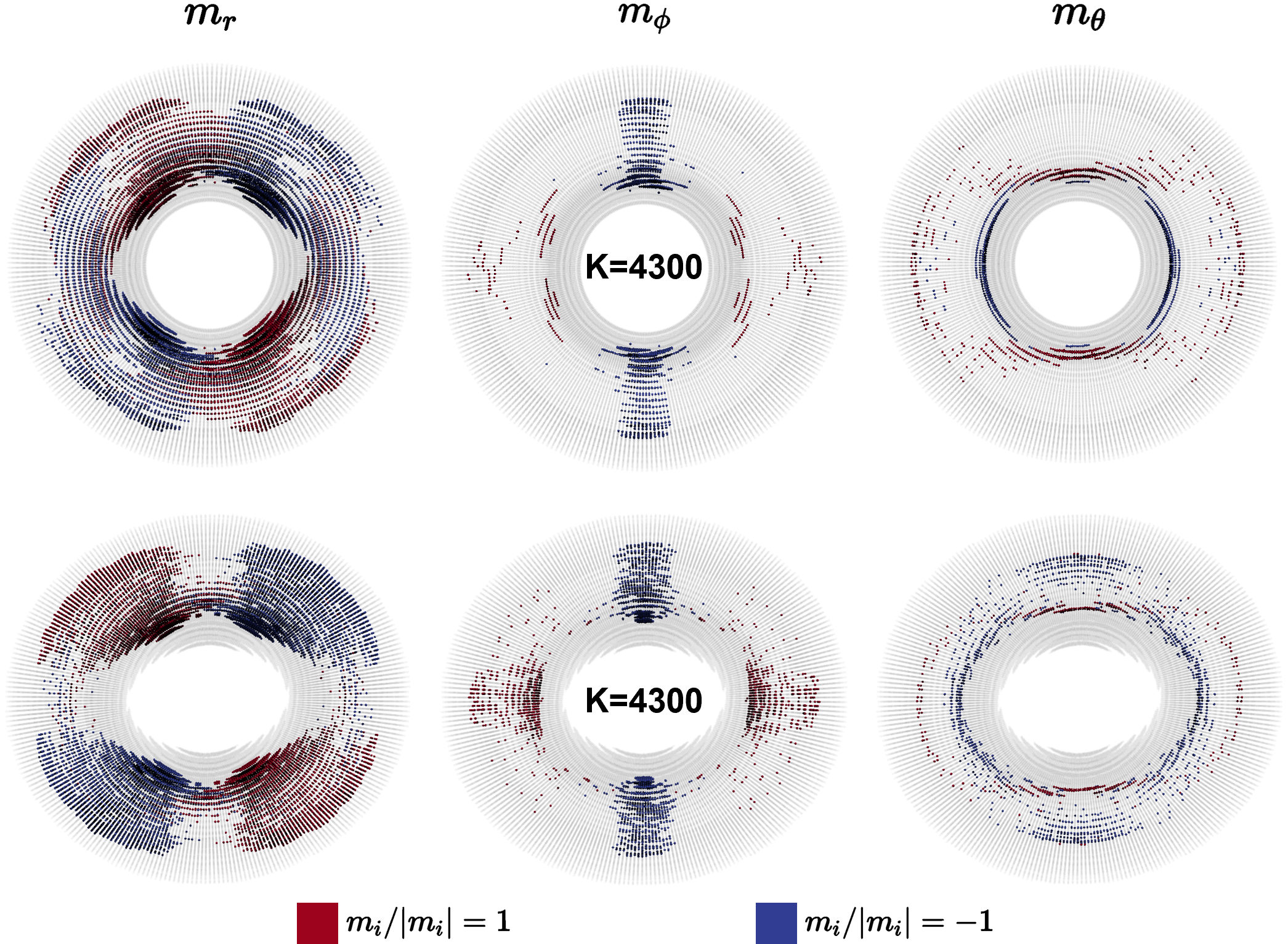}
    \captionsetup{width=\textwidth}
    \caption[magnet configurations of the first 4300 magnets in the radial, toroidal and poloidal directions for uniform grids with circular and plasma shaped cross sections.]{Top view of the magnet configurations in the radial toroidal and poloidal directions for the first 4300 magnets placed in the uniform grid with circular cross section (top) and the equilibrium shaped cross section with the same $V_{eff}$ (bot). The magnetization is normalized.
    Note that the radial direction is by far the most populated one and that the positions close to the VV are preferred over farther away spots. This indicates that the equilibrium cross section is more effective for radially oriented magnets close to the VV, but is less efficient when the magnets have to be placed farther away.}
    \label{fig:UniformGridResultsK=4300}
    \vspace{-3mm}
\end{figure}

The closest layers of magnets to the VV are filled first for both grid layouts, meaning they lead to the biggest reduction of the cost function. Furthermore, the chosen grid spots and corresponding magnet orientations are very similar for the two cases in \cref{fig:UniformGridResultsK=4300}, with radially oriented magnets being the most prevalent. This means that, for a thin layer, using a magnet grid that follows the magnetic surfaces is advantageous. However, as the distance to the plasma increases, the magnets placed on the circular grid have a larger effect leading to a later saturation of the objective function and a lower minimum. 

\begin{figure}
    \centering 
    \includegraphics[width =0.8\textwidth]{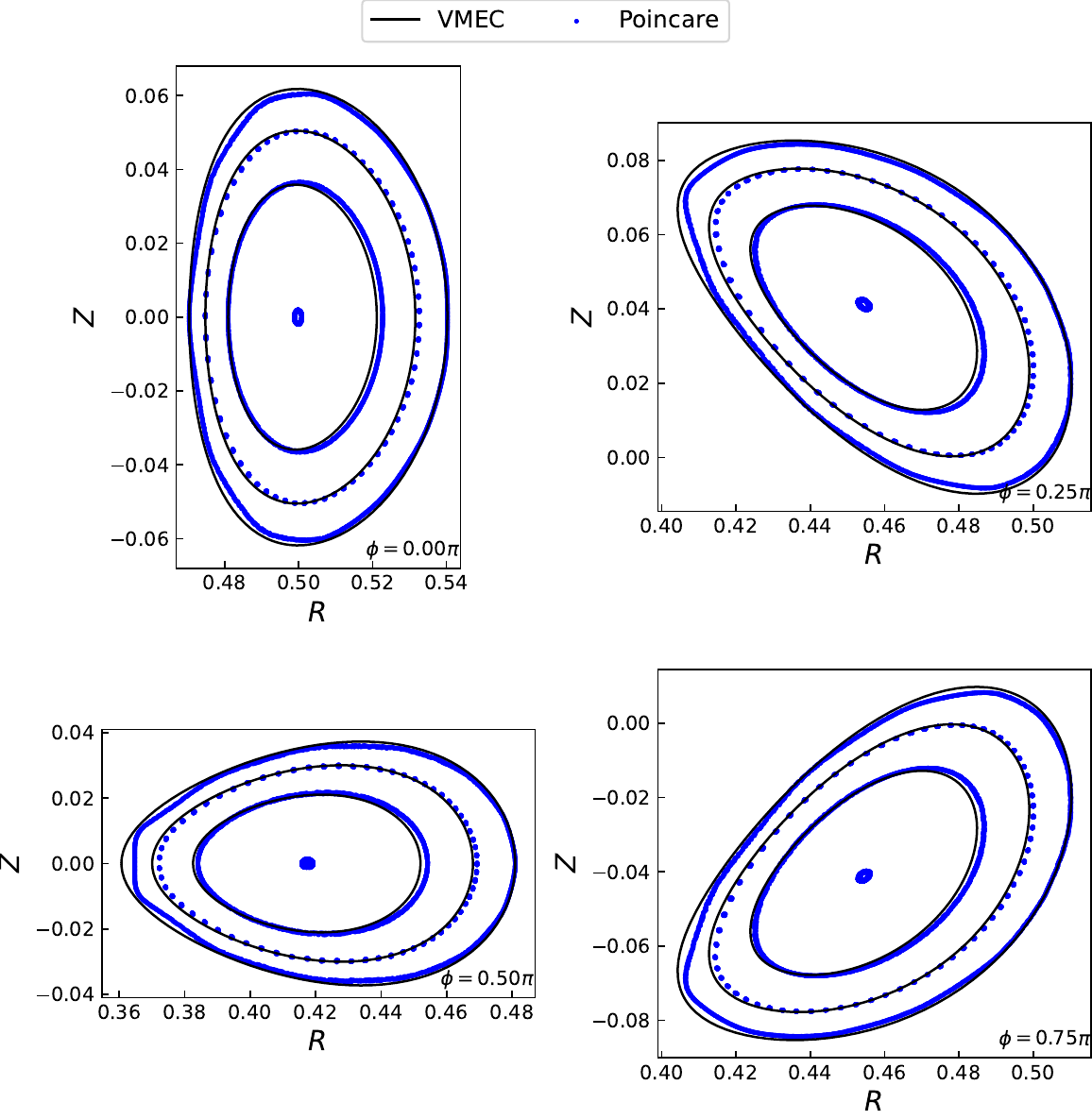}
    \captionsetup{width=\textwidth}
    \caption{Poincaré Plot calculated for the circular uniform grid solution. The target magnetic field surfaces (black) are superimposed with the calculated field lines (blue).}
    \label{fig:PoincareUniformGrid}
\end{figure}

In the present case, we obtained an objective function of $f_B = \SI{6e-6}{\tesla\squared\meter\squared}$, yielding field errors close to 0.1\%. To assess if such an objective is low enough to obtain the target set of nested flux surfaces, the field lines were traced using the SIMSOPT framework with a tolerance of $10^{-14}$. The Poincaré plot, i.e., the tracing of the magnetic field lines along a cross-section plane, on the resulting field of the toroidal field coils and permanent magnets is shown in \cref{fig:PoincareUniformGrid}. We find that the field lines approximately follow the intended surfaces and that there are no signs of magnetic islands. We note that the uniform meshes considered here do not allow for spacing between magnets for a support structure, nor leave space for ports to access the plasma. These approximations will negatively impact the generated field, thus it is important to find ways of including these aspects while keeping $f_B$ low.

As an aside, we note that for designs where the main goal is the use of a thin layer of magnets, where most of the poloidal field is provided by another source, e.g., to do error field corrections to a set of modular coils, the results above point to the use of a magnet grid with the same shape as the boundary of the nested flux surfaces as a more viable option as it offers a faster decrease of the objective function and follows the shape of the coils more closely than a circular shape, which is favorable for maximum use of the available volume. The grid may even be further restricted to only allow normally oriented magnets for ease of assembly as it is the dominant magnet orientation, similarly to Ref. \cite{Zhu2020}.

\subsection{Cubic Magnets and Polarization Study}
\label{section:Magpie}

Up to now, we have only considered uniform magnet grids which are useful for determining a project's viability and allow qualitative comparisons between distinct situations. However, in order to design a magnet array that can be built, one has to consider spacing for a mounting structure and practical magnet orientations. 
The ideal magnet mesh should minimize the number of unique magnet types both in shape and in polarization while maximizing the use of the available volume and the number of possible magnetization directions. For this purpose, the MAGPIE code \cite{Hammond2022}, which was developed with the objective of aiding the design of practical PM arrays, was used.

MAGPIE currently supports three geometries: quadrilaterally-faced hexahedra (\textit{qhex}), curved bricks (\textit{cbrick}), and trapezoidally-enclosed rectangular prisms (\textit{trec}).
The \textit{qhex} concept is intended for perpendicularly oriented magnets. The \textit{cbrick} concept can fit the volume between two toroidal surfaces better than the \textit{trec} concept. However, the magnets increase in volume with the radial coordinate, thus increasing the number of unique magnets. Furthermore, as a consequence of having a curvature, they do not have any rotational symmetries that can be exploited to increase the number of magnetization directions without increasing the number of unique magnets. Since minimizing the number of unique magnet types and maximizing the number of possible magnetization directions are key aspects in an ideal magnet mesh, the \textit{trec} concept was chosen.  

The advantages of this design can be built upon by considering cubic magnets. In the prism family, cubes have the most degrees of rotational symmetry, thus maximizing the possible magnet orientations without increasing the number of unique magnets. Additionally, picking cubes reduces the degrees of freedom by choosing only 1 spatial dimension instead of up to 3 (length, width, and height). Furthermore, the cubic magnet has the least average dipole approximation error for rectangular-cross-section bars and when compared to other shapes, namely diametric and axial magnets \cite{Petruska2013}. This means the estimation of the magnetic field is as accurate as it could be with the chosen optimization tools. Finally, cubic magnets are easily obtained from any magnet manufacturer.
\begin{figure}
    \centering 
    \includegraphics[width =\textwidth]{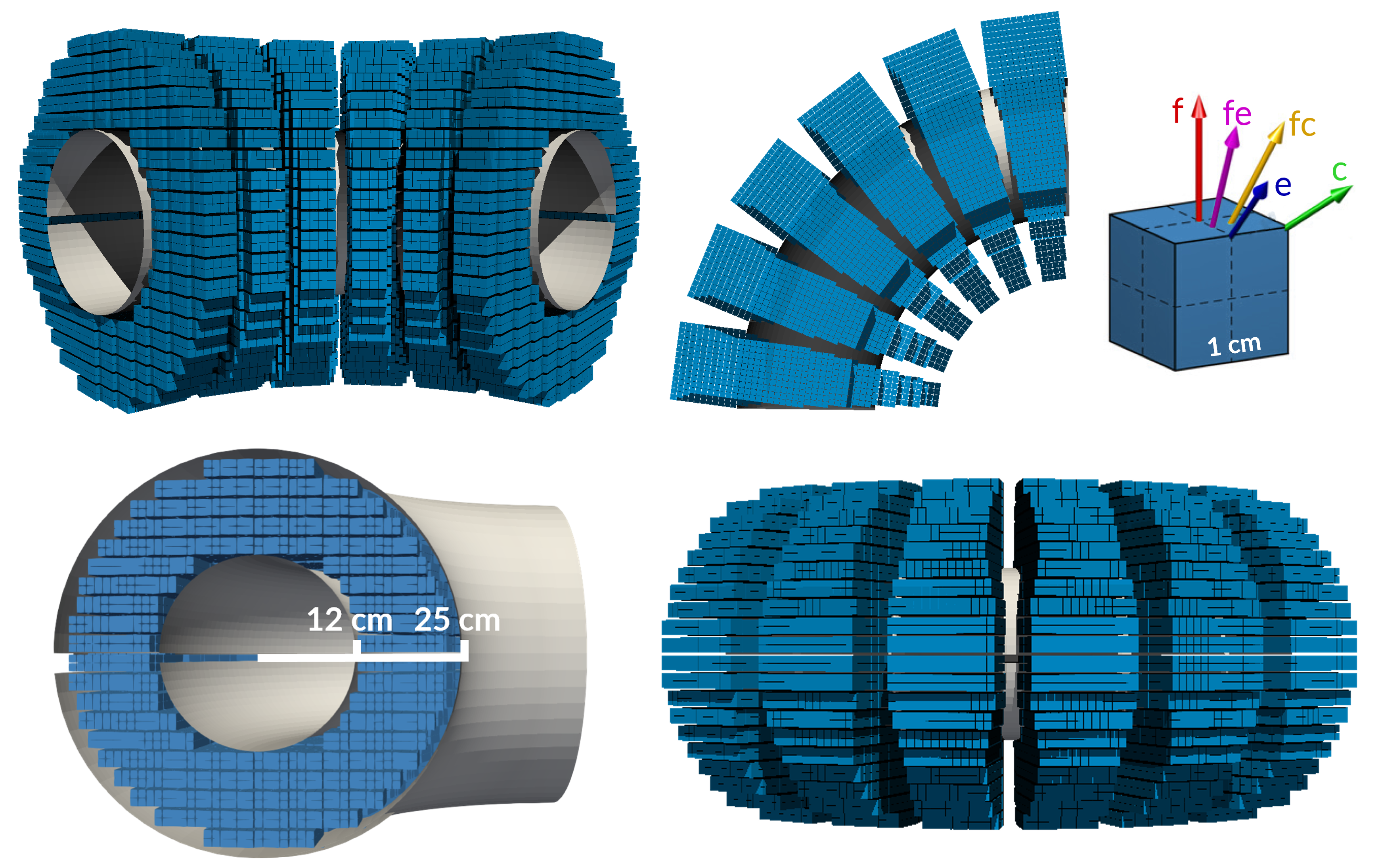}
    \captionsetup{width=\textwidth}
    \caption[Trapezoidally-enclosed rectangular prisms (\textit{trec}) PM array concept with cubic magnets and considered polarization types.]{Trapezoidally-enclosed rectangular prisms (\textit{trec}) PM array design with cubic magnets \cite{Hammond2022}. Top right: considered polarization types in this work \cite{Hammond2022}.}
    \label{fig:RealisticGrid}
    \vspace{-2mm}
\end{figure}

The MAGPIE code enables the specification of spaces between magnets, magnet bundles, and toroidal wedges, accounting for the placement of the mounting structure. Calculating how large these spaces must be requires rigorous structural analysis, determining if the electromagnetic loads are within an allowable stress range. This is beyond the scope of this work. Instead, the required space is overestimated by using the work done for NCSX as reference \cite{Zhu2022,Hammond2022}. NCSX has a major radius of $R_0 = \SI{1.44}{\meter}$ and a minor radius of $a = \SI{0.32}{\meter}$. It is considerably larger than the machine we are considering, hence leading to a conservative approach. In Refs. \cite{Zhu2022,Hammond2022}, magnets within the same bundle are separated by \SI{0.04}{\centi\meter} to allow for a coating layer on each magnet and gluing magnets to one another, the $\SI{6}{\centi\meter}$ bundles/drawers are separated by \SI{0.97}{\centi\meter} and there is a minimum distance of \SI{1.59}{\centi\meter} between toroidal wedges. 

In this work, we consider a $\SI{0.05}{\centi\meter}$ gap between magnets, a $\SI{1}{\centi\meter}$ gap between magnet bundles for each \SI{6}{\centi\meter} of magnetic material (for example, \SI{2}{\centi\meter} drawers are separated by approximately $\SI{0.33}{\centi\meter}$) and the same minimum distance of $\SI{1.59}{\centi\meter}$ between toroidal wedges. The number of toroidal wedges, $N_{tor}$, is picked to match the number of toroidal field coils. This allows the larger portions of the mounting structure to be in between coils, having more freedom in the volume use and easier access to the mounting structure. 
By default we will consider $\SI{1}{\centi\meter}$ magnets and $\SI{2}{\centi\meter}$ bundles (2 rows of magnets per bundle). The magnet array is limited by an inner toroidal surface with a major radius of $R_0 = \SI{0.46}{\meter}$ and a minor radius of $a = \SI{0.12}{\meter}$ and an outer toroidal surface with a major radius of $R_0 = \SI{0.46}{\meter}$ and a minor radius of $a = \SI{0.25}{\meter}$. The MAGPIE inputs and outputs for the arrays used in this work are available in \cite{miguel_madeira_2024_10656980}.

The final PM structure is shown in \cref{fig:RealisticGrid}. This design has a simple and practical mounting structure that facilitates the assembly and disassembly of the magnet mesh (see Refs. \cite{Zhu2022, Hammond2022}), as the magnet array consists of magnet bundles (drawers) that are grouped into a number of toroidal wedges. Within a bundle, the magnets are glued to one another, with the number of magnet layers being chosen as an input. 

Having selected the magnet array design and the spacing for the mounting structure, the possible optimization polarization types are the only aspects left to be defined. 
The following polarization types are considered: face ($f$), corner ($c$), edge ($e$), face-edge ($fe$), and face-corner ($fc$) (see \cref{fig:RealisticGrid}). They have 6, 8, 12, 24, and 24 rotational symmetries, respectively. The magnet price is dependent on the polarization. Polarizations that align with a larger number of faces are harder to manufacture and hence more expensive.

\begin{figure}
    \centering 
    \includegraphics[width =0.5\textwidth,valign=c]{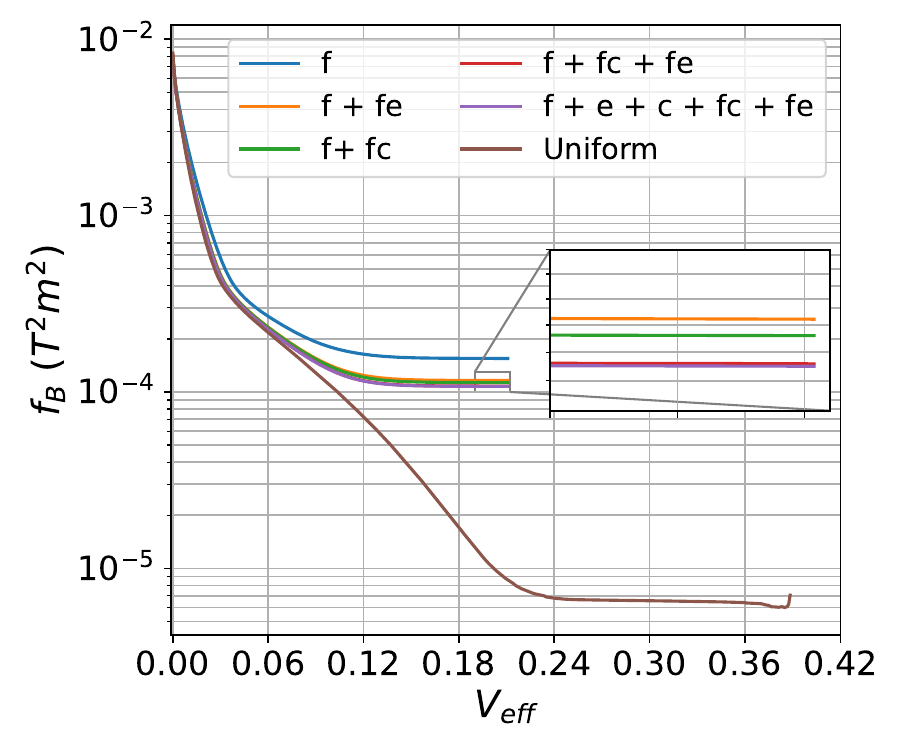}
    \includegraphics[width =0.49\textwidth,valign=c]{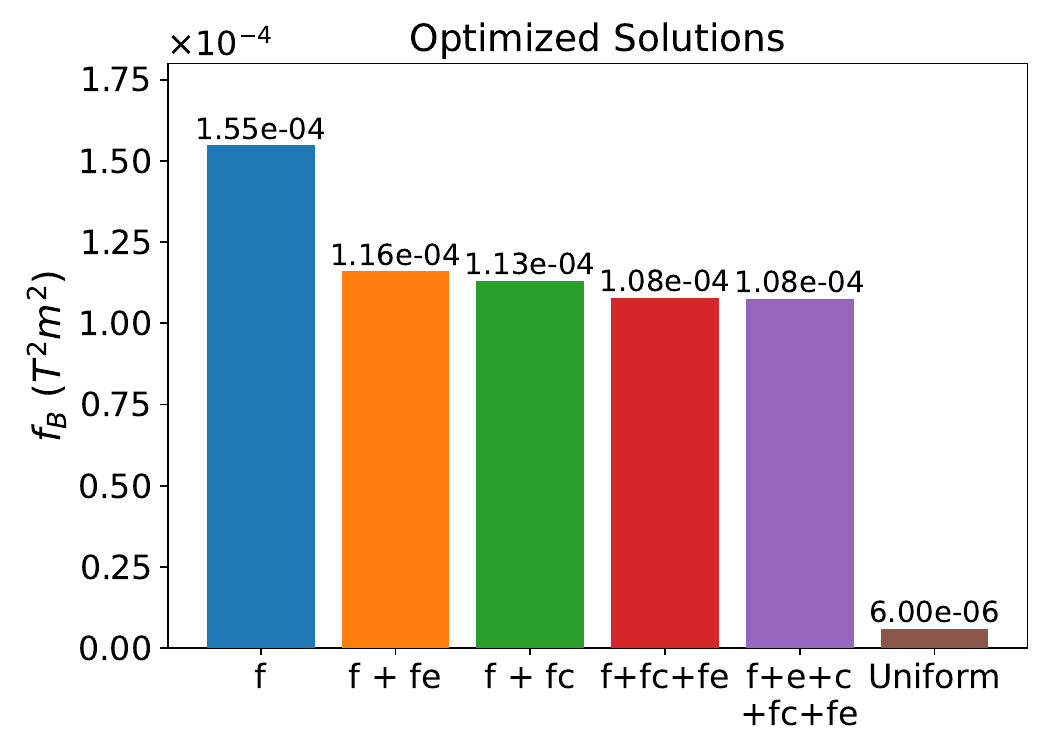}
    \captionsetup{width=\textwidth}
    \vspace{-3mm}
    \caption[Optimization curves and minima for a QA configuration using the ArbVec algorithm with different magnet polarization types.]{Left: Optimization curves for a QA configuration using the ArbVec algorithm with different magnet polarization types. The magnet grid follows the cubic magnet design described in Ref. \cite{Hammond2022} using a circular cross-section. The results are in good agreement with Ref. \cite{Hammond2022}, with the face, face-edge, face-corner combination being the most cost-effective solution. Right: Bar chart of the optimal solution (minimum cost function) for each case.}
    \label{fig:RealisticGridOpt}
\end{figure}

\begin{figure}
    \centering 
    \includegraphics[width =\textwidth]{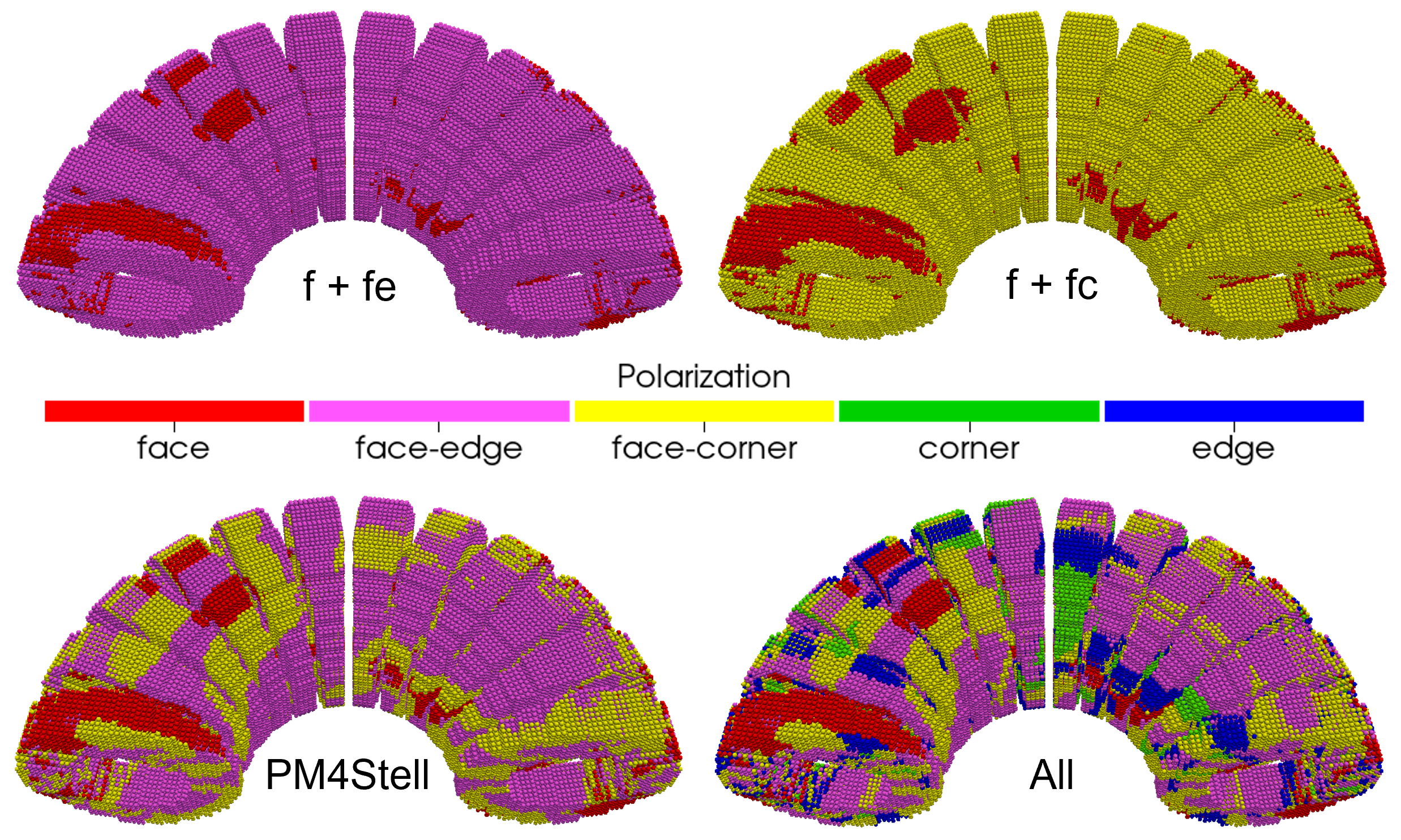}
    \captionsetup{width=\textwidth}
    \vspace{-4mm}
    \caption[Magnet polarizations for the optimal solutions.]{The magnet face (red), face-edge (magenta), face-corner (yellow), corner (green), and edge (blue) polarizations are identified for the optimal solution of the curves in \cref{fig:RealisticGridOpt}. The case with only face-polarized magnets is omitted as all magnets would be red. Note how some patterns remain the same in different solutions, namely the face-polarized magnets always occupy the same spots.}
    \label{fig:RealisticGridConfig}
\end{figure}

In Refs. \cite{Zhu2022, Hammond2022} the $f$, $fe$, $fc$ combination is appointed as the best subset of polarization types. Nonetheless, this result was obtained for an NCSX-like configuration, which is a larger machine with more available magnet volume and a different plasma shape. 
In order to verify these results, the ArbVec GPMO algorithm was used for 5 subsets of magnets. Subsets with edge or corner magnets without $fe$ and/or $fc$ are excluded as the subset with all polarizations is enough to confirm whether these polarizations are advantageous or not. The resulting optimization curves are presented in \cref{fig:RealisticGridOpt} and the resulting magnet configurations in \cref{fig:RealisticGridConfig}, with the relevant statistics being summarized in \cref{tab:RealisticGrid}.

The first remark to make is the large difference in field error between the uniform toroidal grid case and a more realistic design (see \cref{fig:RealisticGridOpt}). It is indeed important to consider the spacing between magnets. 
Secondly, we note that the results are consistent with Ref. \cite{Hammond2022}, as there is close to no difference ($< 0.5\%$) when adding the edge and corner polarizations to the $f$ + $fc$ + $fe$ combination. However, adding these two polarizations would increase the cost of our magnet grid considerably due to the increase in the number of unique magnets and the fact that edge and corner polarizations are more expensive to manufacture. 

\begin{table}
    \caption{Magnet polarization fraction when all array spots are filled, for the 5 considered polarization subsets. The magnet grid has a total of 52512 magnets.}
    \vspace{-2mm}
    \small
    \begin{indented}
    \item[]\begin{tabular}{@{}lllllll}
    \br
    \makecell{Polarization subset} & face & face-edge & face-corner & edge & corner & sum\\
    \mr
    $f$ & 100\% & - & - & - & - & 100\%\\\vspace{1mm}
    $f$ + $fe$ & 7.5\% & 92.5\% & - & - & - & 100\%\\\vspace{1mm}
    $f$ + $fc$ & 12.2\% & - & 87.8\% & - & - & 100\%\\\vspace{1mm}
    \makecell{$f$ + $fe$ + $fc$ (PM4Stell)}& 7.3\% & 58.4\% & 34.3\% & - & - & 100\%\\\vspace{1mm}
    \makecell{$f$ + $fe$ + $fc$ + $e$ + $c$} & 7.2\% & 44.8\% & 23.7\% & 14.4\% & 9.9\% & 100\%\\
    \br
    \end{tabular}
    \end{indented}
    \vspace{-6mm}
    \label{tab:RealisticGrid}
\end{table}

In Ref. \cite{Hammond2022}, the $fe$ and $fc$ polarizations were not considered individually. However, from \cref{fig:RealisticGridOpt} it is possible to see that the $f$ + $fc$ combination is more effective for PM optimization than the $f$ + $fe$ combination. In contrast, when PM4Stell-like or all polarizations are considered, face-edge magnets are the most prevalent and in the $f$~+~$fe$ and $f$~+~$fc$ cases, the face-edge magnets are more dominant over the face-centered magnets than the face-corner magnets (see \cref{tab:RealisticGrid} and \cref{fig:RealisticGridConfig}). If there is some leeway in a design with PM4Stell-like polarizations, it may be worth considering dropping the face-edge magnets to reduce the number of unique magnets. 

From \cref{fig:RealisticGridConfig}, it is possible to notice that the face-polarized magnets always occupy the same spots, forming a similar red pattern independently of the other polarizations involved, even though they are the least prevalent magnets in all cases where multiple polarizations are involved. The number of face magnets drops only $0.2\%$ from the $f$+$fe$ to the $f$+$fe$+$fc$ cases and then $0.1\%$ to the situation with all considered polarizations. This suggests they are more effective in specific cells, but less adaptable, possibly due to their lower number of rotation symmetries. On the other hand, $fe$ and $fc$ lose a large portion of their spots when edge and corner magnets are introduced, yet this barely impacts the objective function. 
In the optimization with all possible polarizations, the fraction of occupied spots follows the same order as the number of rotation symmetries ($f < c < e < fc = fe$) evidencing the importance of choosing magnet geometries that allow for multiple rotation symmetries. 

Given the findings in this study, we will adopt the face, face-corner, and face-edge polarizations for the remainder of this work, due to their cost-effectiveness. 

\section{From ISTTOK to ISTELL}
\label{section:ISTELL}

While the results obtained in the previous sections already took ISTTOK's geometry into account, some aspects of the machine still need to be considered to obtain a comprehensive design, namely the VV minor radius, the shift between the VV and the TF coils, and the volume occupied by the plasma accessing ports. In the following sections, these aspects will be addressed. The coil currents are fixed to match ISTTOK's operating toroidal magnetic field of $\SI{0.5}{\tesla}$. 
Due to its small size and long lifetime, ISTTOK is a prime candidate for transforming a tokamak into a stellarator. Retaining its naming scheme, we refer to its stellarator counterpart as ISTELL for short. First, a new magnetic equilibrium that aligns with the device’s characteristics is optimized. Subsequently, an initial permanent magnet optimization study is conducted, excluding consideration of any diagnostic ports. The diagnostic ports are then modeled and taken into account in the following optimization. Lastly, a scenario in which the vacuum vessel is replaced is considered. 

\subsection{Equilibrium Optimization}\hfill
\label{section:equilibrium}

\noindent
The first step in converting ISTTOK to a stellarator is finding a suitable magnetic equilibrium. Having a predetermined VV greatly increases the difficulty of adapting existing equilibria to our needs. Instead, we optimize a magnetic equilibrium. The main optimization objectives for ISTELL's equilibrium are, in order of significance, fitting ISTTOK's VV, having a high degree of quasisymmetry, and being well described by the near-axis expansion \cite{Landreman2019,Jorge2020}. The second target serves to have better particle confinement and the third to allow experimentally validating near-axis codes that have been recently developed, as well as to simplify future analytical studies. The equilibrium was chosen to have $\mathrm{nfp} = 2$ and to be stellarator symmetric. 

With these objectives in mind, the optimization of ISTELL's equilibrium \cite{miguel_madeira_2024_10656980} was performed using the following cost function, $J$,
\begin{eqnarray}
    \fl J =  w_0\sum_{i = 1}^{N_C} d_{i} + w_1\sum_{s_j} \left \langle \left(\frac{1}{B^3}[(N - \iota M)\textbf{B}\times\nabla B\cdot \nabla \psi - (MG + NI)\textbf{B}\cdot\nabla B ]\right) \right \rangle \nonumber\\
    + w_2(\iota - 0.122) ^ 2 + w_3(A - 10.7)^2 ,
    \label{eq:equilibriumobjective}
\end{eqnarray}
with
\begin{equation}
 d_{i} = \int_{C_i} \int_{S} \max(0, d_{\min} - \| \mathbf{r}_i - \mathbf{s} \|_2)^2 ~dl_i ~ds ,
\end{equation} and $N = 0$.
The parameter $s$ is the toroidal magnetic flux enclosed by a flux surface, normalized to the flux at the boundary. The first term in \cref{eq:equilibriumobjective} is a threshold on the distance ($\| \mathbf{r}_i - \mathbf{s} \|_2$) between $N_C$ curves, $C_i$, and the last closed surface, $S$. The curves are evenly distributed circumferences with radius $r_{C_i} = \SI{8.5}{\centi\meter}$ that model ISTTOK’s VV. The second term \cite{Landreman2021} optimizes for quasi-axisymmetry. The third term defines a target for the rotational transform, $\iota$, so the solution does not converge to the axisymmetric case and reduces the plasma shaping when compared to the previously used QA equilibrium. The fourth term defines a target aspect ratio to attempt to find a solution with as high a plasma volume as possible. Nonetheless, the $\iota$ and $A$ values are greatly influenced by the need for the equilibrium to fit ISTTOK's VV. The number of Fourier modes that define the boundary was progressively increased up to a maximum of 3, while $d_{\min}$ was progressively decreased down to a minimum of \SI{0.6}{\milli\meter}. This strategy had the most success in guaranteeing the equilibrium fit the VV. The optimization is formulated using the SIMSOPT framework  \cite{SIMSOPT}, with each magnetic equilibrium being calculated through the VMEC code \cite{Hirshman1983}. A local minima of the cost function is found using the Levenberg-Marquardt algorithm, which is SciPy's default algorithm, with an absolute step of $10^{-7}$ and a relative step of $10^{-5}$.

\begin{figure}
    \centering
    \includegraphics[width = 0.49\textwidth, valign=c]{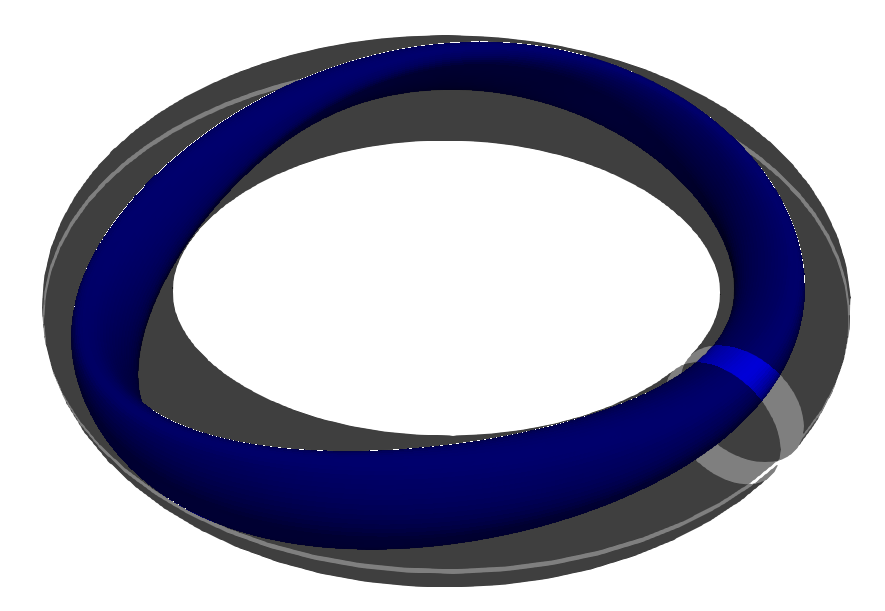}
    \includegraphics[width = 0.43\textwidth, valign=c]{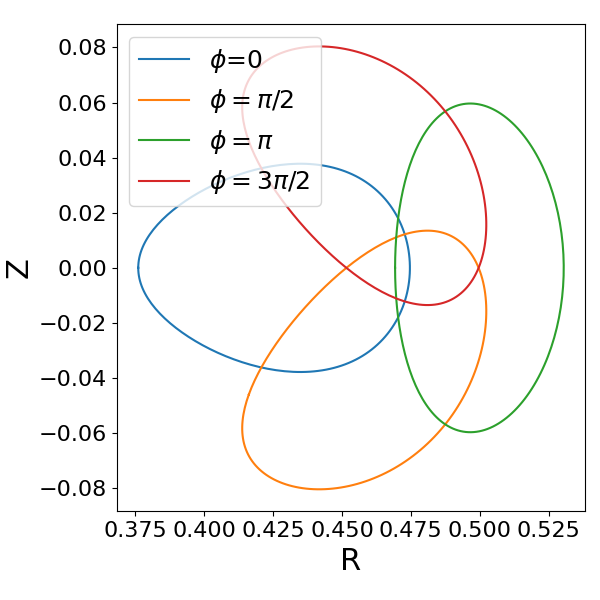}
    \vspace{-6mm}
    \caption[ISTELL's equilibrium fits ISTTOK's vacuum vessel.]{Left: 3D plot of the last closed surface for the obtained QA equilibrium (blue) inside ISTTOK's VV (low opacity black). Right: ISTELL's equilibrium poloidal cross sections.}
    \label{fig:ISTELLfits}
\end{figure}
\begin{figure*}[t]
    \centering
    \includegraphics[trim={0 0 0 0.4cm}, clip, width = 0.45\textwidth, valign=c]{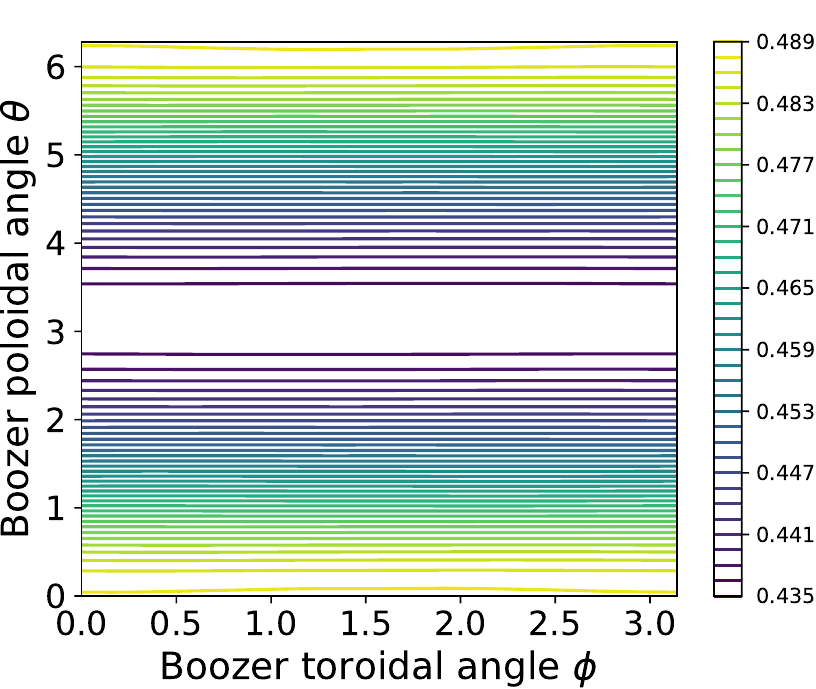}
    \hspace{2mm}
    \includegraphics[trim={0 0 0 1cm},clip, width = 0.49\textwidth, valign=c]{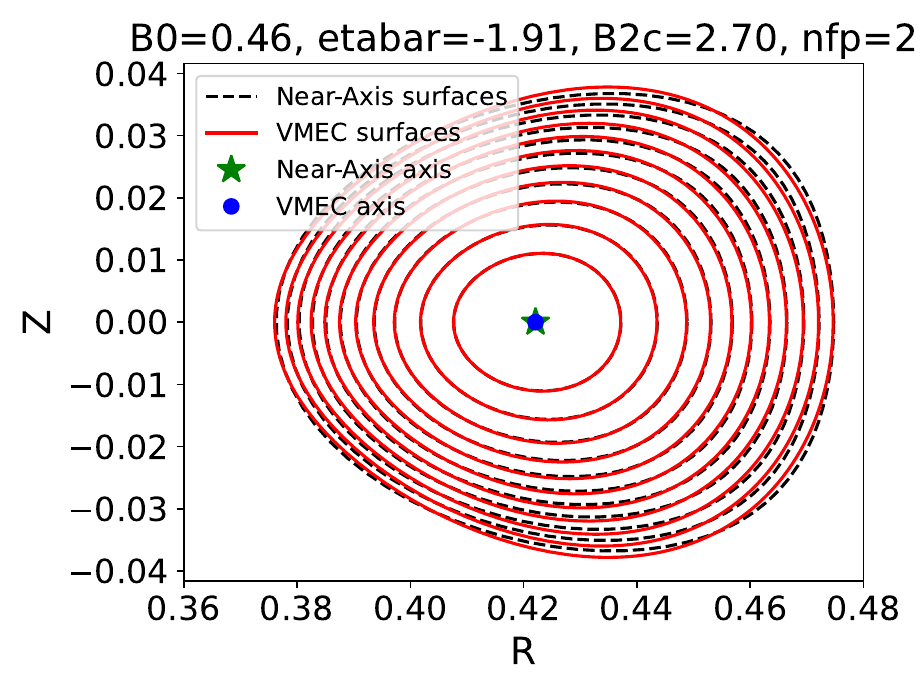}
    \vspace{-2mm}
    \caption{Left: the quasisymmetry of the found QA magnetic equilibrium is demonstrated by the straight contours of the field strength B on the $s = 0.5$ flux surface in the $\theta$-$\phi$ plane in Boozer coordinates. The contours were calculated using the Booz\_XFORM code \cite{Sanchez2000}. Right: Superimposition of the target magnetic surfaces and the magnetic surfaces calculated with the near-axis expansion after the fitting of the parameters to the VMEC magnetic field.}
    \label{fig:ISTELLQSNA}
    \vspace{-4mm}
\end{figure*}

The resulting equilibrium fits ISTTOK's vessel, \cref{fig:ISTELLfits}, has an aspect ratio of $A = 10.8$, an almost constant rotational transform of $\iota = 0.122$ and a quasisymmetry residual of $3.49\times 10^{-7}$. The high degree of quasisymmetry is evident in \cref{fig:ISTELLQSNA} (left), with the contours of $B$ being straight in Boozer coordinates. This guarantees that collisional transport would almost certainly be weaker than turbulent transport. The equilibrium's axis shape can be described as 
\begin{eqnarray}
    R_0(\phi)[m] &= 0.4572 - 0.0368\cos(2\phi) + 0.0018\cos(4\phi),\\
    z_0(\phi)[m] &= -0.03667\sin(2\phi) + 0.0018\sin(4\phi),
\end{eqnarray} 
corresponding to two field periods. For the first and second-order near-axis parameters, we find that $\overline{\eta} = \SI{-1.91}{\per\meter}$ and $B_{2c} = \SI{2.70}{\tesla\per\meter\squared}$. The parameters were obtained by fitting a near-axis solution to the VMEC magnetic field. The parameters $\sigma(0)$ and $B_{2s}$
were set to zero so the configuration is stellarator symmetric. We also choose $I_2 = 0$ and
$p_2 = 0$ so the configuration is a vacuum field. The resulting configuration has $\iota_0 = 0.125$, and the boundary
shape for $A = 10.8$ is shown in \cref{fig:ISTELLfits}.
From \cref{fig:ISTELLQSNA} we can also verify that the near-axis expansion can, as intended, fit the equilibrium, although it was optimized with VMEC.

%\subsection{Permanenent Magnet Optimization}

\subsection{ISTELL Shifted and Aligned Scenarios}
\label{section:ISTELLscenarios}

To have a higher average magnetic field, ISTTOK's coil and VV axes have a $\SI{6}{\centi\meter}$ shift between the two axes (see \cref{tab:ISTTOK} and \cref{fig:ISTTOKshifted}). 
In this shifted configuration, there is little available space for permanent magnet placement on the high field side. On the other hand, the total available volume is larger than in an aligned case, as the volume increases with the radius. It is possible to align the coil and VV axis, increasing the available volume on the high field side, by pushing the coils to the inside of the machine. However, due to ISTTOK's large number of coils, $N = 24$, and limited space on the inner side of the torus, this requires reducing the number of coils to $N \leq 17$. 
\begin{figure}
    \centering 
    \includegraphics[width =0.5\textwidth]{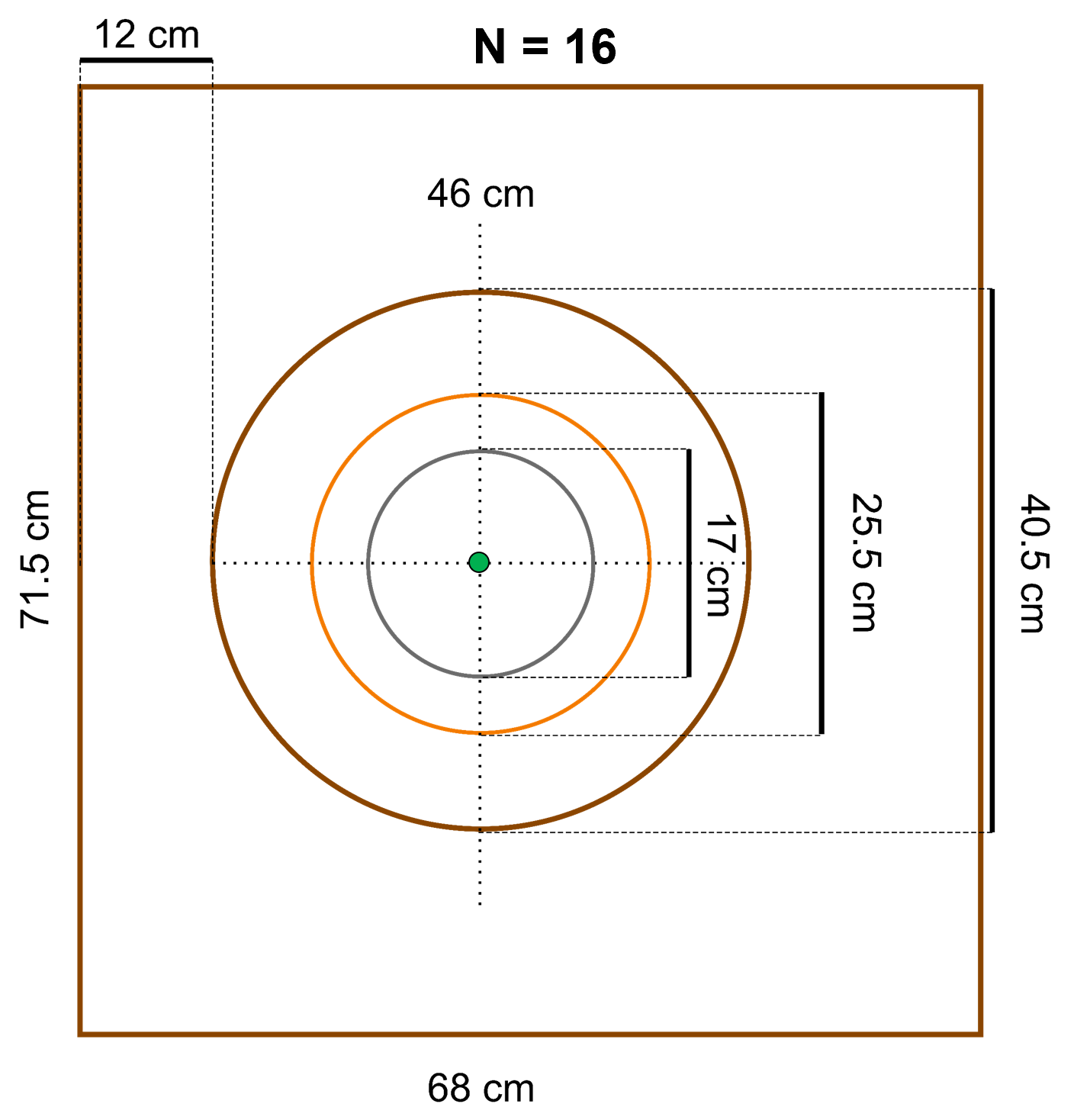}
    \captionsetup{width=\textwidth}
    \vspace{-4mm}
    \caption[Scheme of ISTELL's aligned scenario.]{Cross section of the proposed aligned layout that reduces the number of toroidal field coils from $N = 24$ to $N = 16$ to increase the available space on the high field side. The system's dimensions are identified, namely the toroidal field coils are represented in brown, the widest section of the copper shell is represented in orange and the narrowest section of the vessel is represented in grey.}
    \label{fig:ISTTOKShift}
    \vspace{-4mm}
\end{figure} 

In order to evaluate which setup is better, PM optimizations were performed with the existing shifted arrangement and a new proposed layout, \cref{fig:ISTTOKShift}, with the two axes aligned and $N = 16$ coils (corresponding to 4 coils per half field period). A reduction in the number of coils increases the coil ripple. Nonetheless, this plays directly into the advantages of employing a permanent magnet grid and can likely be therefore compensated. 

For both cases, the inner limiting surface for the magnets is axisymmetric with the same axis as the VV and a $\SI{12.75}{\centi\meter}$ radius. This corresponds to the VV external radius at its thickest section of $\SI{11.25}{\centi\meter}$ plus the $\SI{1.5}{\centi\meter}$ copper shell. The outer radius of the VV is $\SI{9.4}{\centi\meter}$ for the majority of the machine, so this is a rather risk-conservative approach. The considered dimensions are schematized in \cref{fig:ISTTOKshifted} and \cref{fig:ISTTOKShift}. A magnet side of $dr = \SI{1}{\centi\meter}$ was considered. In line with the concepts from \cref{section:Magpie}, for both layouts, the possible magnet orientations conform to the face, face-edge, and face-corner polarizations (see Ref. \cite{Zhu2022}), and the number of toroidal subdivisions is the same as the number of considered coils, $N_{tor} = N$. The coils were modeled as circular lines with the same current that produce $\SI{0.5}{\tesla}$ on the VV axis.

MAGPIE assumes that the VV is centered in the permanent magnet grid. To build the shifted mesh, a grid with a larger radial extent was used as a starting point. Subsequently, the magnets outside a toroidal surface centered at the coil axis with a $\SI{20}{\centi\meter}$ radius were removed. The coils are circular and have a $\SI{20.25}{\centi\meter}$ radius, therefore this yields $\SI{0.5}{\centi\meter}$ of leeway. For the aligned case, no additional steps were required other than running MAGPIE with the adjusted inputs to the system.

\begin{figure}
    \vspace{-2mm}
    \centering 
    \includegraphics[width =0.6\columnwidth]{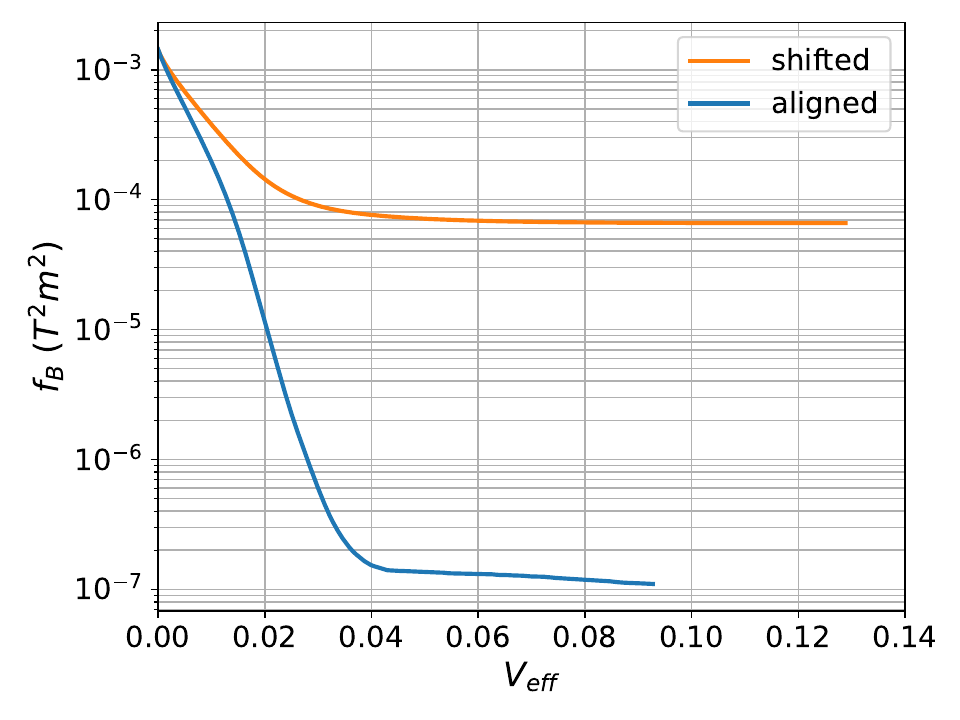}
    \vspace{-3mm}
    \caption{ Optimization curves for a QA equilibrium using the ArbVec algorithm for two configurations where the coil and vacuum vessel axes are either shifted or aligned.}
    \label{fig:ShiftedVSAlignedOptCurves}
    %\vspace{-5mm}
\end{figure}

Applying the ArbVec GPMO algorithm to the described designs, the optimization curves from \cref{fig:ShiftedVSAlignedOptCurves} were obtained. From these results, it is clear that the aligned case is advantageous when compared to the shifted case showing an $f_B$ that is roughly 3 orders of magnitude lower.
\begin{figure}
    \centering 
    %\vspace{-2mm}
    \includegraphics[width = 0.95\columnwidth]{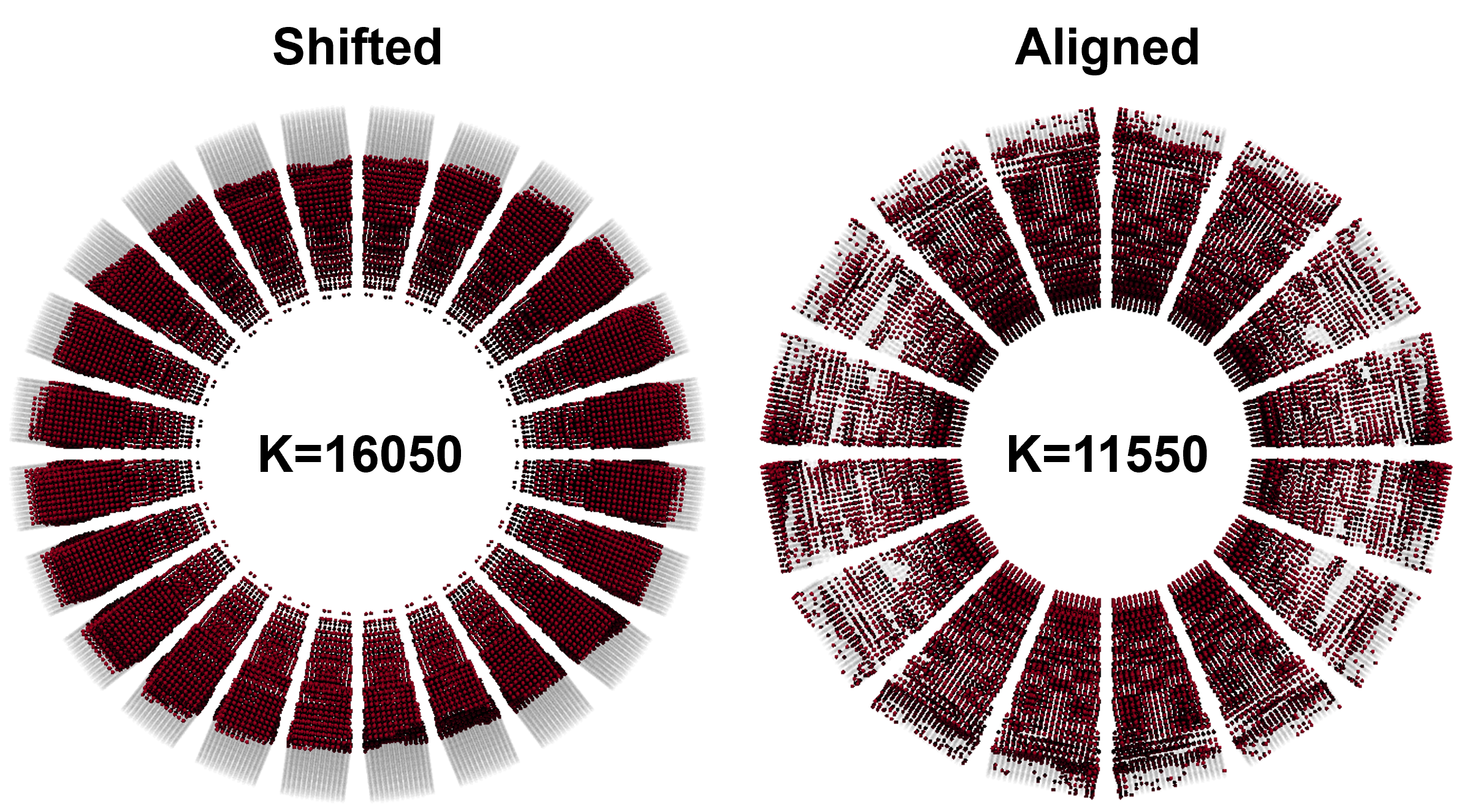}
    \vspace{-2mm}
    \caption{Magnet configurations for 50\% occupied effective volume for the shifted and aligned layouts. The low field side spots on the shifted scenario are mostly unoccupied suggesting they are not effective at reproducing the magnetic equilibrium. The aligned scenario shows a more even distribution of the magnets.}
    \label{fig:ShiftedVSAlignedMagnetPlacement}
\end{figure} 

Examining the magnet placements for the two scenarios (\cref{fig:ShiftedVSAlignedMagnetPlacement}) when the objective function, $f_B$, plateaus, provides insight into the advantages of the aligned layout. The shifted layout shows a clear concentration of magnets closer to the VV and the spots on the low field side are unoccupied. For the aligned layout, although these spots are still preferred a more uniform distribution of the magnets is observed. Since the advantage of the shifted layout is having more available space on the low field side and the positions closer to the VV and on the high field side are preferred, it is only natural that the aligned mesh produces better results. 

\begin{figure}
    \centering 
    \includegraphics[width = 0.95\textwidth]{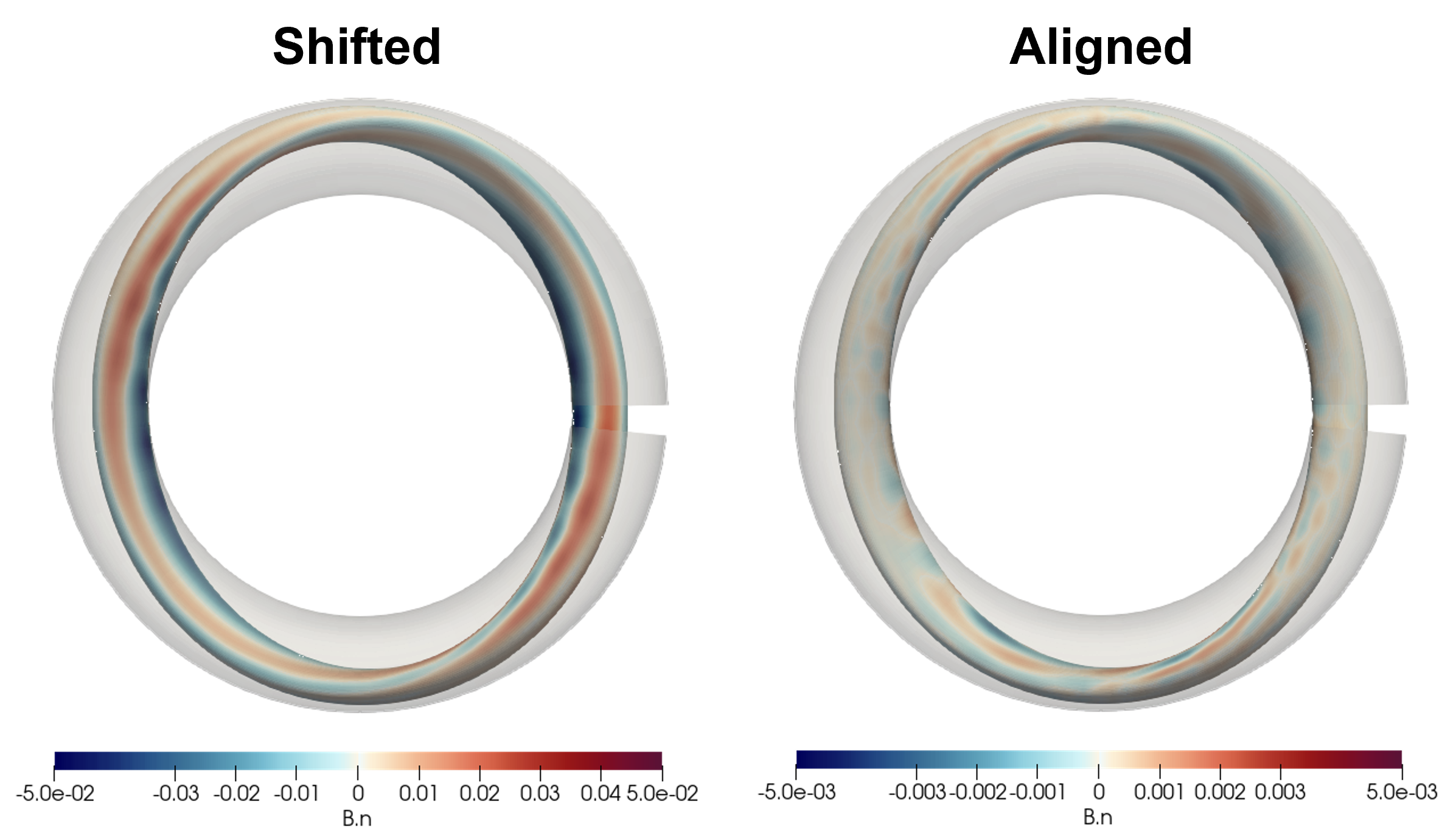}
    \captionsetup{width=\textwidth}
    \caption[$\textbf{B}\cdot \textbf{n}$ on the equilibrium surface for the shifted and aligned cases.]{Field error, $\textbf{B}\cdot \textbf{n}$, on the equilibrium surface for the shifted (left) and aligned cases (right). The VV is also represented in low-opacity grey. Note that the $\textbf{B}\cdot \textbf{n}$ scale is one order of magnitude lower in the aligned case. Furthermore, the maxima in the aligned case are more localized.
    For the shifted case, the maximum is present on the high field side where the magnetic equilibrium is closest to the VV, due to the lack of magnetic material in this region. For the aligned case, the maximum occurs where the equilibrium is farther away from the VV due to the larger distance between the magnets and the plasma.}
    \label{fig:ShiftedVSAlignedBn}
\end{figure}

The respective weaknesses of both configurations are further evidenced by the $\textbf{B}\cdot \textbf{n}$ color maps of the optimal solution represented in \cref{fig:ShiftedVSAlignedBn}. For the shifted case, the maximum of $\textbf{B}\cdot \textbf{n}$ is present on the high field side where the magnetic equilibrium is closest to the VV. This results from a lack of magnetic material in this region due to the low available volume on the high field side.
For the aligned case, the maximum occurs on the high field side where the magnetic equilibrium is farther away from the inside of the VV. This results from the larger distance between the magnets and the plasma. 

\subsection{Plasma Accessing Ports}
\label{section:ports}

The precise calculation of crucial parameters such as electron temperature and density is key to the success of magnetic confinement experiments. These parameters not only serve as essential inputs for fusion and plasma physics codes but also play a pivotal role in motivating future experimental studies and influencing the design of upcoming machines.  As such, we will now take into account the fact that ISTTOK has a total of 33 plasma-accessing ports that serve multiple purposes.

The modeling of these ports becomes imperative for PM optimization as they reduce the available volume, which negatively impacts the attainable field quality. Furthermore, since they were designed for a tokamak, ISTTOK's ports do not follow stellarator or field period symmetry. As GPMO leverages such symmetries to reduce computation time, this greatly increases the computational cost of the design, as it requires the use of the full magnet array. For this reason, the magnet size was increased from $\SI{1}{\centi\meter}$ to $\SI{2}{\centi\meter}$ effectively reducing the free variables by a factor of 8.

\begin{figure}
    \centering 
    \includegraphics[width =\linewidth]{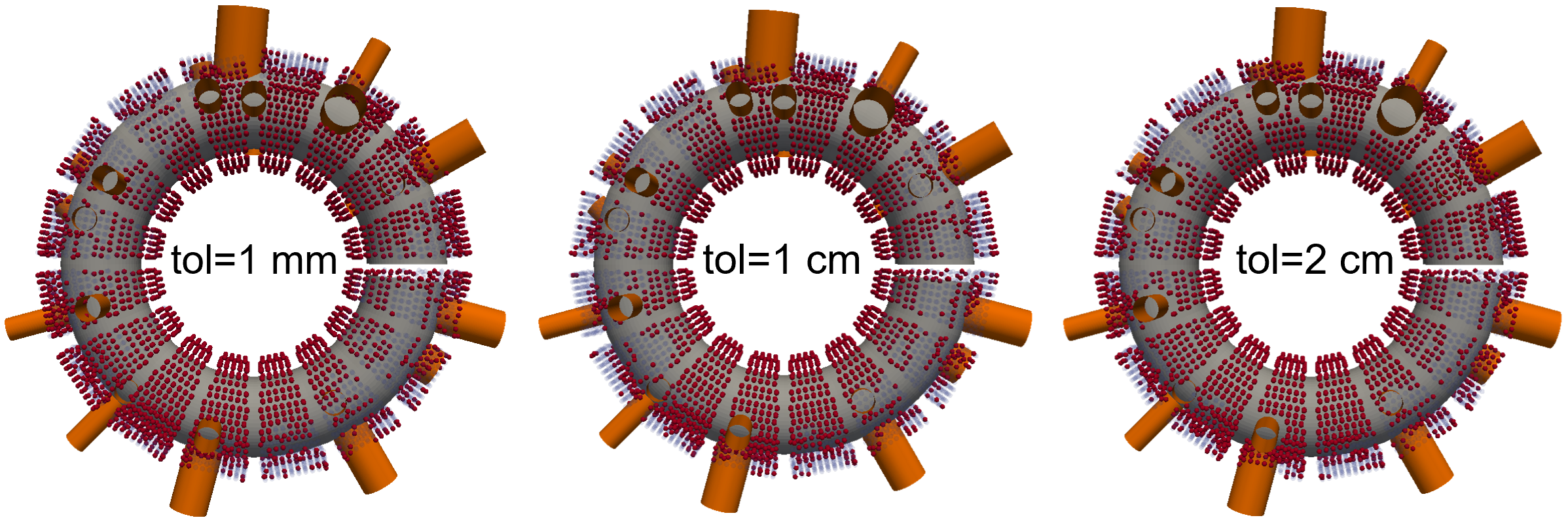}
    \vspace{-5mm}
    \caption{Magnet placement (red) for the $tol = \SI{1}{\milli\meter}$ GPMOb, $tol = \SI{1}{\centi\meter}$ and $tol = \SI{2}{\centi\meter}$ once the plateau has already been reached at $V_{eff} = \SI{0.04}{\meter\cubed}$. The ports are represented in orange.}
    \label{fig:ISTELLports}
    \vspace{-3mm}
\end{figure}

The ports, illustrated in \cref{fig:ISTTOKscheme} and \cref{fig:ISTELLports}, were modeled as cylinders, with the radius given by one of three standard flange sizes, DN40, DN60 and DN100. The height of the cylinders was considered equal to the coil radius (see \cref{tab:ISTTOK}) for the ports in use and the port height plus the thickness of the standard flange sizes for the remaining ports. This means that only the currently used ports in ISTTOK would remain functional and the remaining ports would be rendered unusable in exchange for a larger available volume for magnet placement. To include ports in the magnet mesh, each PM is selectively removed if assessed to be inside the volume of the cylinders modeling the ports.

\begin{figure}
    \centering 
    \includegraphics[width =0.5\textwidth, valign=c]{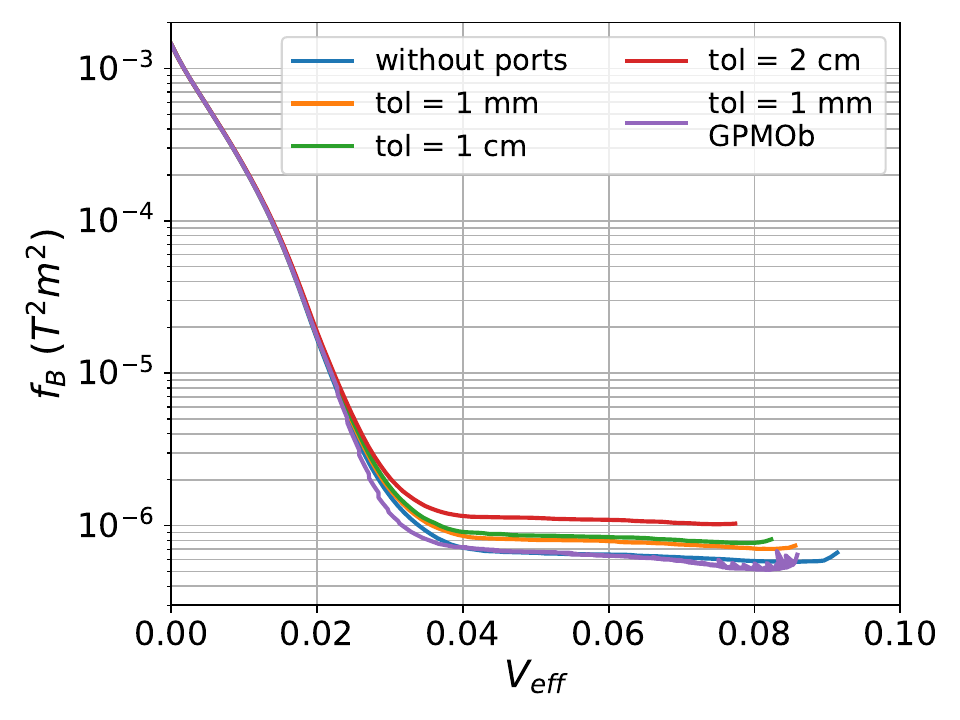}
    \includegraphics[width =0.49\textwidth, valign=c]{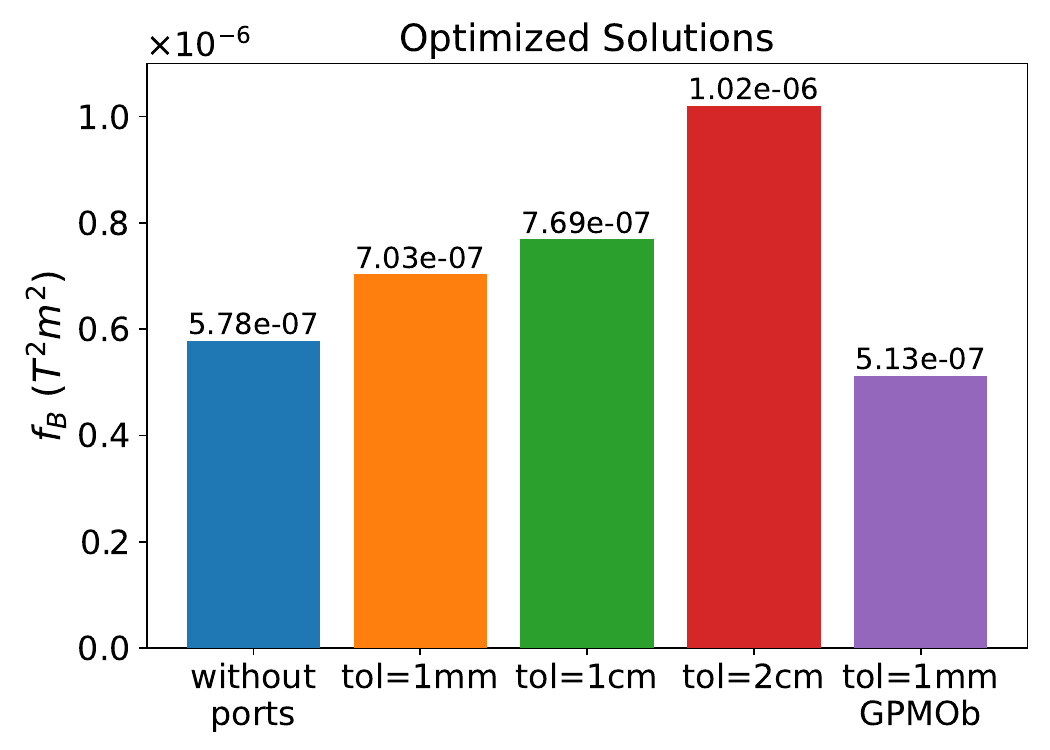}
    \captionsetup{width=\textwidth}
    \vspace{-4mm}
    \caption[PM Optimization curves and minima for ISTELL's QA configuration using the ArbVec algorithm with diagnostic ports considering different tolerances.]{Left: PM Optimization curves for ISTELL's QA configuration using the ArbVec algorithm for a full \SI{2}{\centi\meter} magnet grid with diagnostic ports considering different tolerances. The $dr=\SI{2}{\centi\meter}$ case without ports is plotted for reference. For the curve in purple, the optimization algorithm was upgraded to the backtracking version (GPMOb). Right: Bar chart of the optimal solution for each case. Introducing diagnostics increases the cost function by at most $22\%$ and it remains below $10^{-6}$ as long as the considered tolerances are reasonably low. The downside of introducing ports is lower than the upside of using the backtracking algorithm.}
    \label{fig:ISTELLdiag}
    \vspace{-1mm}
\end{figure}

We represent a tolerance, $tol$, i.e., an increased distance around the ports, both in radius and in height, to guarantee a risk-averse design.
With this in mind three grids were considered with $tol = \SI{1}{\milli\meter}$, $tol = \SI{1}{\centi\meter}$ and $tol = \SI{2}{\centi\meter}$ in radius and height. The three grids were optimized with the ArbVec GPMO algorithm. The $tol = \SI{1}{\milli\meter}$ mesh was also optimized with the backtracking ArbVec GPMO algorithm considering six adjacent magnets (the first neighbors) and a threshold angle, i.e., the angle between two magnet orientations after which they are removed, of $5\pi/6$. The resulting optimization curves are represented in \cref{fig:ISTELLdiag} and the magnet configurations for the solution at a $V_{eff}$ of $\SI{0.04}{\meter\cubed}$ are represented in \cref{fig:ISTELLports}.

\begin{figure}
    \centering 
    \includegraphics[width =\textwidth, valign=c]{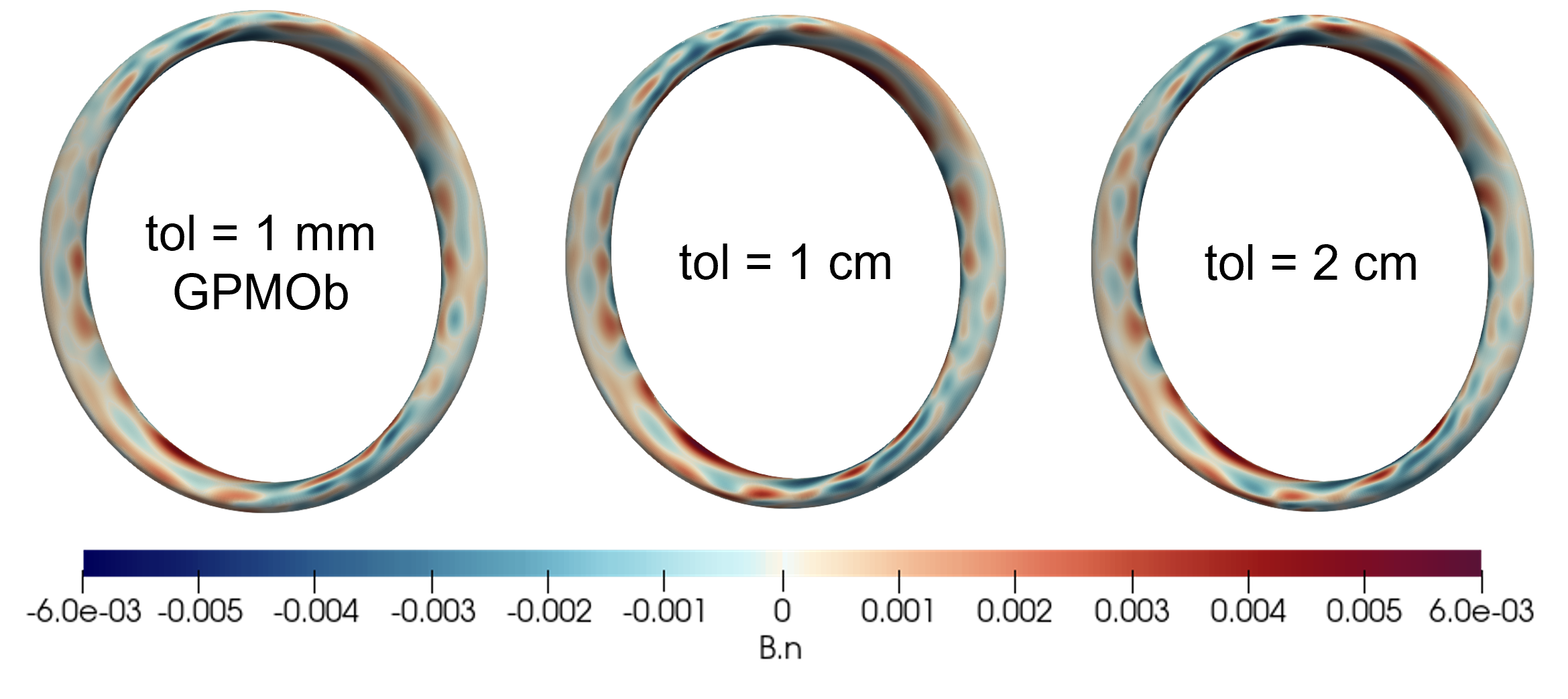}
    \captionsetup{width=\textwidth}
    \vspace{-5mm}
    \caption{Error field, $B\cdot n$, color-map viewed from the top for the backtracking, $tol = \SI{1}{\centi\meter}$ and $tol = \SI{2}{\centi\meter}$ cases. }
    \label{fig:ISTELLdiagBn}
    \vspace{-2mm}
\end{figure}

The introduction of diagnostic ports increases the cost function by at most $22\%$, with $f_B$ remaining below $10^{-6}\SI{}{\tesla\squared\meter\squared}$. In fact, the reduction of the error field from upgrading to the backtracking algorithm is superior to this increase. We find that even with a grid that is not stellarator or field period symmetric, it is still possible to find magnet configurations that can reproduce an intended field that has both symmetries. Furthermore, there is some freedom left when setting tolerances around the ports. With $\SI{1}{\centi\meter}$ of leeway there is an increase of $10\%$ compared to the $\SI{1}{\milli\meter}$ case. Note that the radius of a DN40 flange is $r_\text{DN40} = \SI{3.5}{\centi\meter}$, so a $\SI{1}{\centi\meter}$ tolerance is already substantial. 
The average error field, $\langle \mathbf{B}\cdot \mathbf{n} \rangle$, is $0.18\%$, $0.16\%$, $0.20\%$ and $0.22\%$ for the $dr = \SI{2}{\centi\meter}$ with no ports, $tol = \SI{1}{\milli\meter}$ backtracking, $tol = \SI{1}{\centi\meter}$ and $tol = \SI{2}{\centi\meter}$ situations, respectively. A color-map of the field error is represented in \cref{fig:ISTELLdiagBn} for the backtracking, $tol = \SI{1}{\centi\meter}$ and $tol = \SI{2}{\centi\meter}$ cases. 

\begin{figure}
    \centering 
    \includegraphics[width =\textwidth, valign=c]{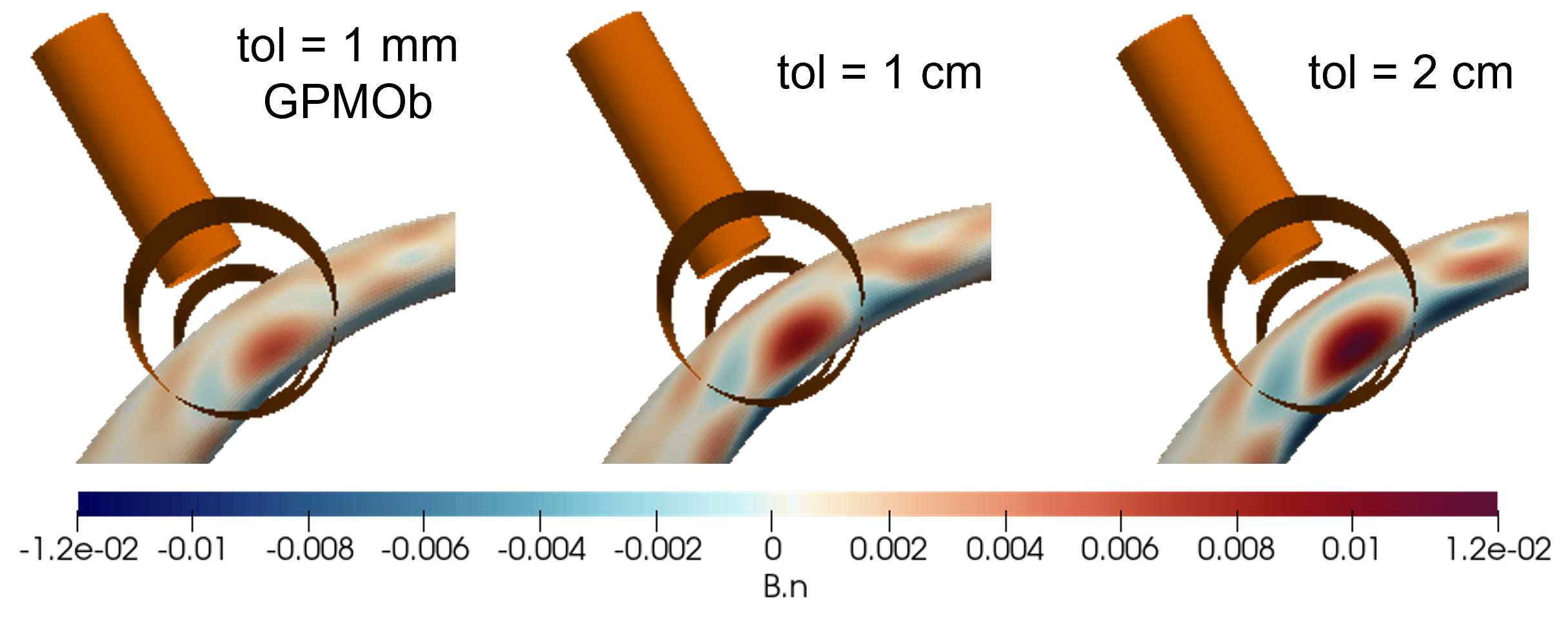}
    \captionsetup{width=\textwidth}
    \vspace{-2mm}
    \caption[Close up of $B\cdot n$ for the region where we have the largest ports.]{Close up of the $B\cdot n$ color-map for the region where we have the largest ports (orange), seen from the bottom. From left to right, for the backtracking, $tol = \SI{1}{\centi\meter}$ and $tol = \SI{2}{\centi\meter}$ cases. As the tolerance around the ports increases so does the maximum error field.}
    \label{fig:ISTELLdiagBnmaxima}
    \vspace{-3mm}
\end{figure}

Although the average error field does not increase substantially with the introduction of ports, the maximum values of the error field tend to increase. This effect is more prevalent as the tolerance around the ports is increased. This is evident for the region with the largest ports, which is also responsible for the largest field error (see \cref{fig:ISTELLdiagBnmaxima}). Defining reasonable tolerances and finding strategies for correcting the field error around ports could be key in the use of permanent magnets in magnetic confinement devices.

\begin{figure}
    \vspace{-2mm}
    \centering 
    \includegraphics[width =\textwidth]{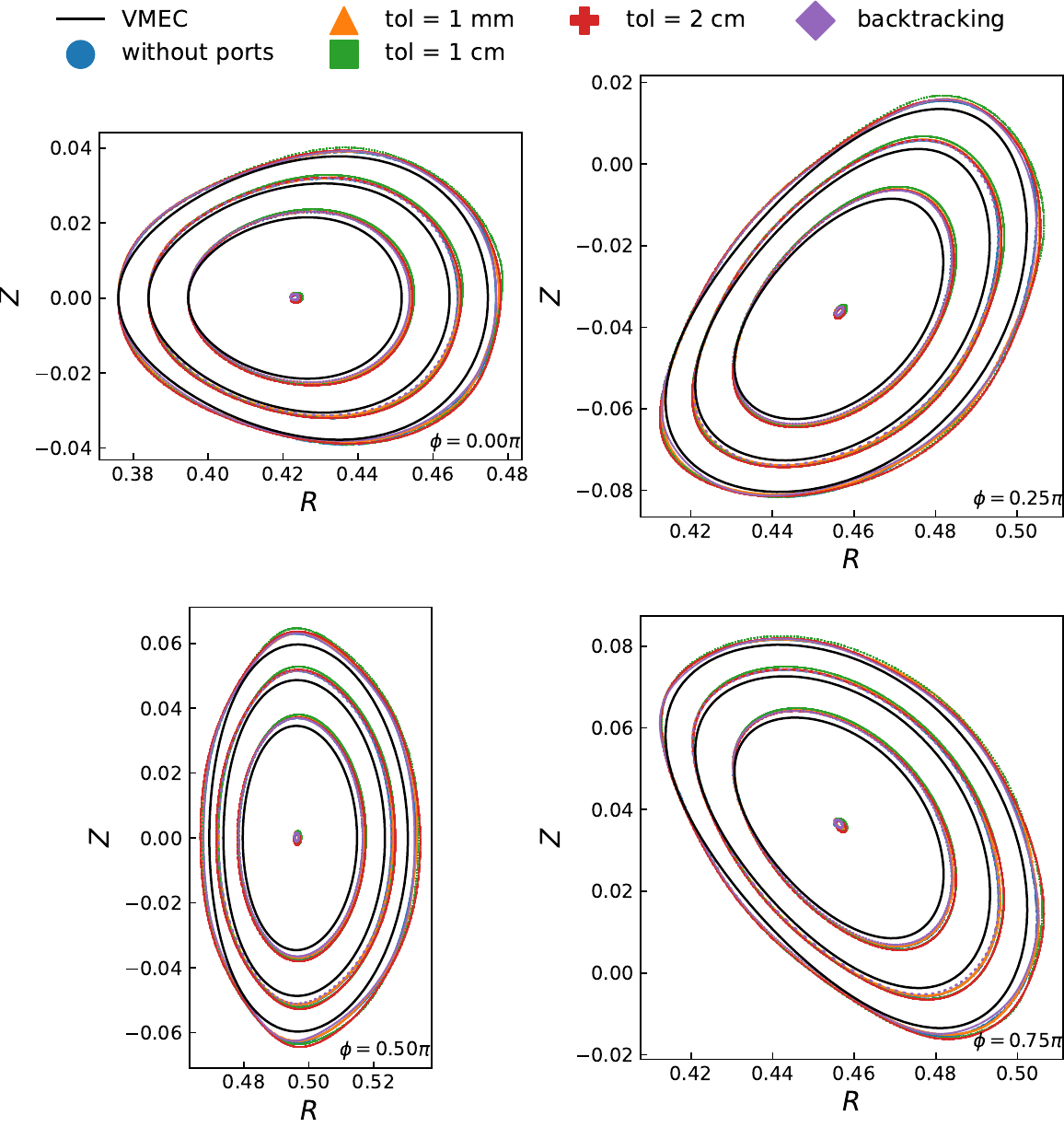}
    \vspace{-7mm}
    \caption{Poincaré Plot calculated using the SIMSOPT framework with a tolerance of $10^{-14}$ for the optimized solutions of the ISTELL equilibrium for a magnet mesh with $\SI{2}{\centi\meter}$ cubes and diagnostic ports at different tolerances. Even for the $tol = \SI{2}{\centi\meter}$ model the VMEC equilibrium is nicely followed and there are no magnetic islands.}
    \label{fig:poincareISTELLdiag}
    \vspace{-3mm}
\end{figure}

To confirm the quality of the fields produced by the obtained permanent magnet configurations, the Poincaré plots for the found solutions are shown in \cref{fig:poincareISTELLdiag}.
The field lines in the Poincaré plots follow the intended magnetic equilibrium approximately for all cases. Furthermore, there are no detected magnetic islands. The field inaccuracy predicted by $f_B$ corresponds to an outward expansion of the magnetic field lines that is also more prevalent in the cases with higher $f_B$, namely the $\SI{2}{\centi\meter}$ tolerance model. The Poincaré plots here presented greatly support the viability of the ISTELL design using a precise QA equilibrium. Furthermore, the solution could be further improved by using smaller magnets and by reducing the gaps between magnets and magnet groups, although the latter entails a careful structural analysis. 
In the absence of ports, the reduction in $f_B$ between the scenarios with $\SI{1}{\centi\meter}$ and $\SI{2}{\centi\meter}$ cubic magnets is approximately $417\%$. Consequently, a comparable enhancement is anticipated even when the ports are considered.

\subsection{Different Vacuum Vessel}
\label{section:diffVV}

In \cref{section:VVshape}, it was concluded that a VV with the same cross-section as the equilibrium posed no significant advantage when compared to a circular VV. Therefore, the natural choice would be to keep ISTTOK's VV. However, for equilibrium optimization, it is considerably easier to choose the VV \textit{a posteriori}. Even if a circular vessel is to be used, a scan of the variation of the objective function with the minor and major radius can be performed. Furthermore, considering a different vessel enables designing the plasma accessing ports so that the stellarator and field period symmetries are kept, thus reducing the number of optimization variables and the computation time. Finally, obtaining a new VV would solve ISTTOK's current shifted configuration difficulty of not having enough space on the high field side, with the added benefits of increasing the available volume for magnets when compared to the aligned case and keeping the 24 toroidal field coils.

For this purpose, we perform an equilibrium optimization allowing for the possibility of using a different VV. We recover the QA equilibrium used in \cref{chapter:designchoices} rescaled to $R_0 = \SI{0.52}{\meter}$, the coil axis' radius. While such configuration would not fit ISTTOK's VV, it has a higher rotational transform, $\iota = 0.167$, as well as a lower aspect ratio, $A = 9.3$, and a lower quasisymmetry residual, $\SI{9.1e-8}{}$. 
Two magnet grids are considered, one that follows the equilibrium cross-section with the inner surface $\SI{1}{\centi\meter}$ away from the last closed surface of the plasma, and one that has a circular cross-section with a radius of $\SI{12.5}{\centi\meter}$. The magnet arrays follow MAGPIE's \textit{trec} design. The spacing between magnets follows the previously defined estimation. The magnets have $dr=\SI{1}{\centi\meter}$ and face, face-edge, or face-corner polarization. We revert back to using 24 coils to produce $B_0 = \SI{0.5}{\tesla}$ and therefore, $N_{tor} = 24$ groups of magnets.

\begin{figure}
    \centering 
    \includegraphics[width =0.49\textwidth, valign=c]{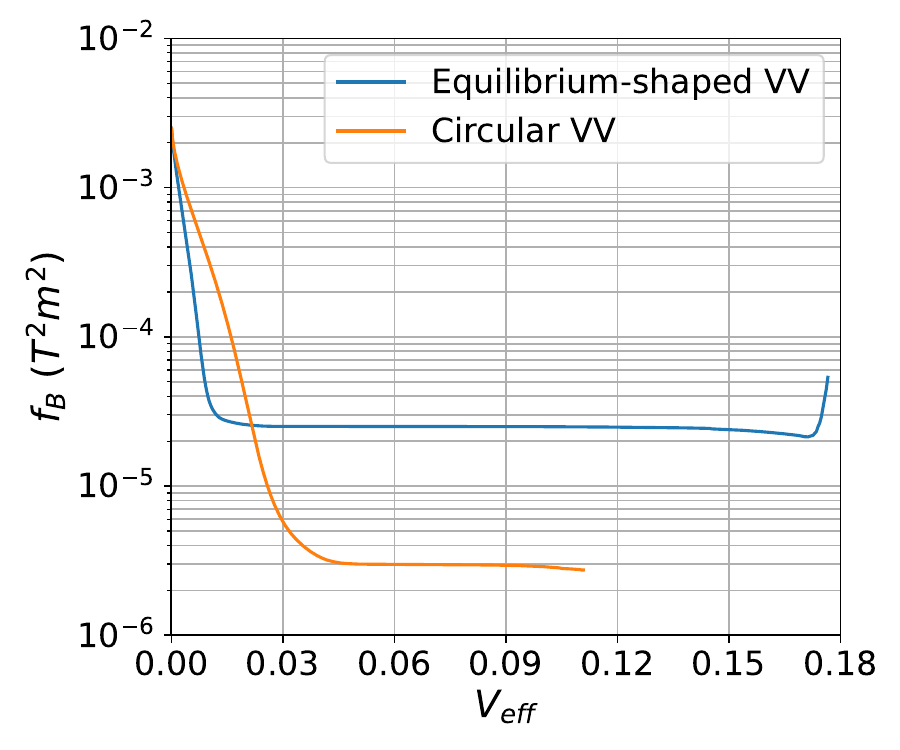}
    \includegraphics[width =0.49\textwidth, valign=c]{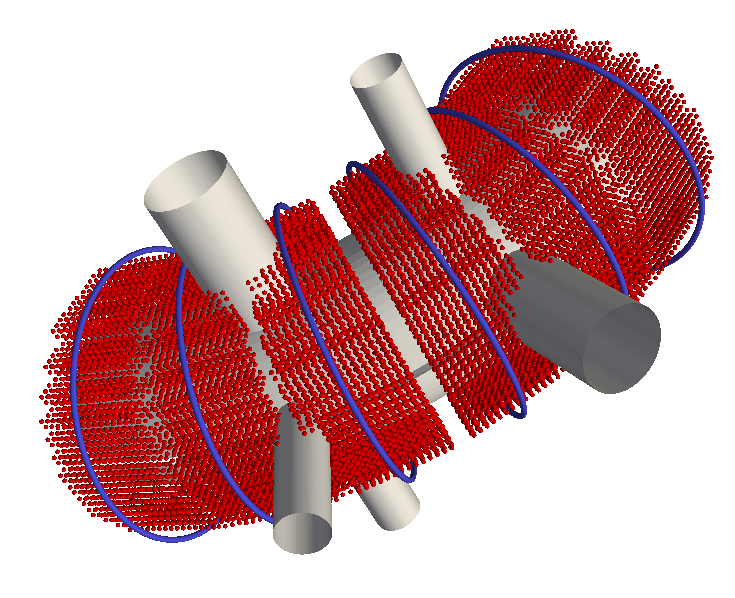}
    \captionsetup{width=\textwidth}
    \vspace{-3mm}
    \caption{Left: PM optimization curves for a QA equilibrium using the ArbVec GPMO algorithm in two scenarios where ISTTOK's VV is replaced - one where the cross-section is circular and another where it follows the last closed surface.  Right: Isometric view for a half field period of the magnet grid (red) for the circular VV with ports (grey) and ISTTOK's circular toroidal field coils (blue).}
    \vspace{-4mm}
    \label{fig:ISTELLdiffVV}
\end{figure}

To retain a similar number of ports to the ones in use for ISTTOK, a total of 20 ports (5 per half field period ) were considered. The port standards were also picked to be as close in number to ISTTOK as possible. The ports were placed between toroidal wedges to minimize the number of removed dipoles. The ports, as well as the magnet grid for a half-field period, are represented in \cref{fig:ISTELLdiffVV}.

The magnet placements were optimized using the baseline ArbVec GPMO algorithm and the corresponding optimization curves are plotted in \cref{fig:ISTELLdiffVV}. The plasma-shaped vessel shows $f_B > 10^{-5}$, considerably larger than the results obtained in the previous section. However, the circular vessel is more promising, having a minimum objective function of  $f_B = \SI{2.74e-6}{\tesla\squared\meter\squared}$, despite its less efficient use of space leading to a lower available volume for magnet placement. This further corroborates the claims from the uniform toroidal grid study that keeping a circular VV is advantageous for magnet optimization, even when considering a more realistic grid with spacing for a support structure and well-thought orientations.

\begin{figure}
    \centering 
    \includegraphics[width =0.8\textwidth]{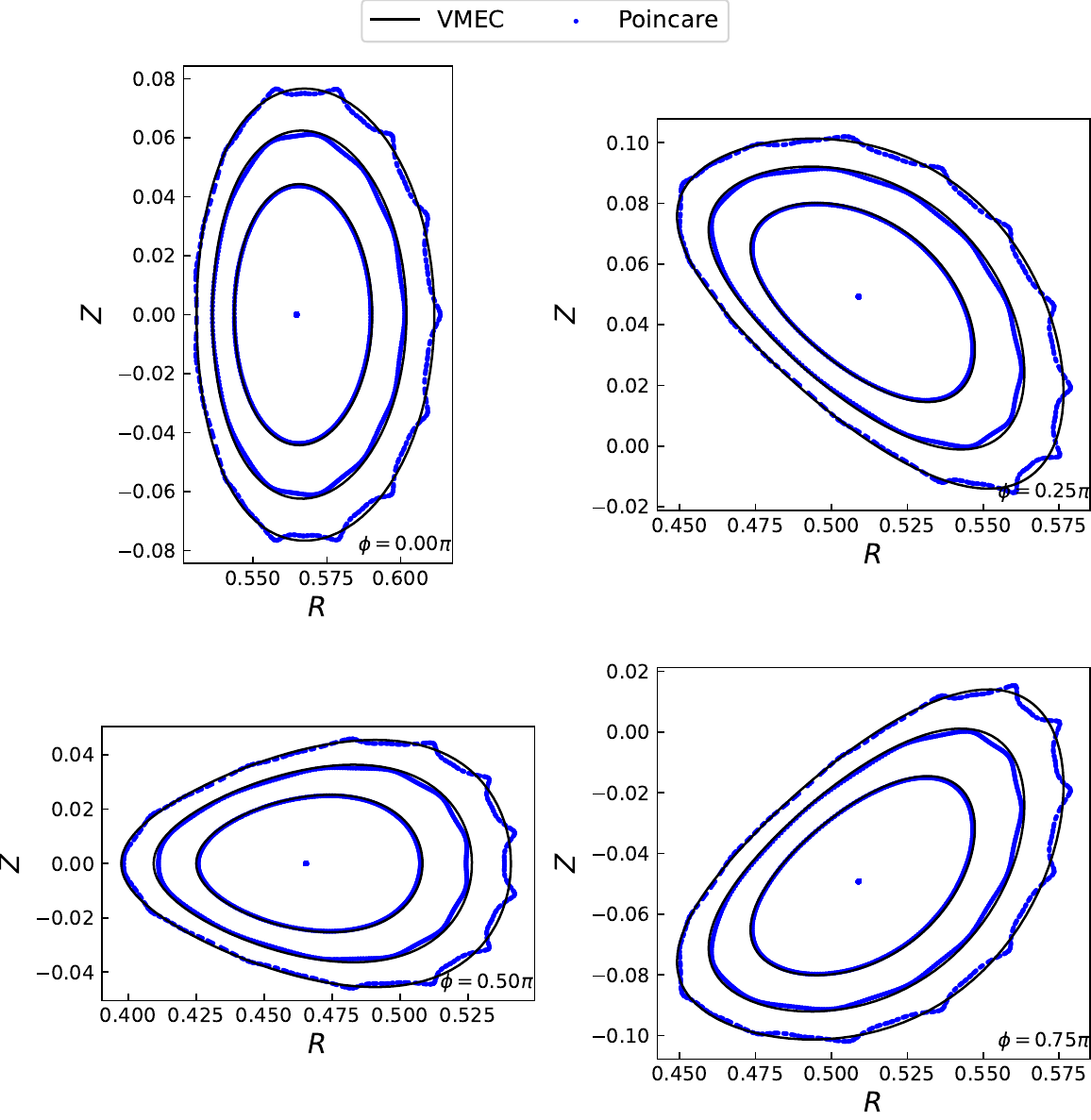}
    \captionsetup{width=\textwidth}
    \vspace{-4mm}
    \caption[Poincaré plots  for ISTELL's different vacuum vessel QA scenario.]{Poincaré plot calculated using the SIMSOPT framework with a tolerance of $10^{-14}$ of the optimized solution for a QA equilibrium using a larger circular vessel than ISTTOK's with 20 ports. Although the magnetic field lines tend to follow the VMEC equilibrium, at the last represented surface, the beginning of 12 magnetic islands are visible indicating there is a resonance with the 2/12 mode.}
    \label{fig:PoincarediffVV}
    \vspace{-1mm}
\end{figure}

To further evaluate the field quality, the Poincaré plot for the circular case was calculated (see \cref{fig:PoincarediffVV}). We see that the field lines closely follow the intended magnetic surfaces. For the last closed surface, a resonance with the 2/12 mode is observed, resulting in 12 magnetic islands. As an aside, we note that these islands could be used for the design of an island divertor, a concept that has an efficient heat load distribution, advantageous scaling with input power, and good impurity screening \cite{Schmitz2021, Jakubowski2021}. A deeper understanding of island divertors could play a pivotal role in developing viable fusion reactors. Therefore, a compact magnetic confinement device that can run a large number of experiments and where design modifications are easily implemented could be very attractive. 

If economically viable, switching ISTTOK's VV for another with a circular cross-section, $R_0 = \SI{0.52}{\meter}$, with stellarator and field period symmetric ports and the minor radius to be defined by the chosen magnetic equilibrium would make ISTELL considerably easier from a design standpoint. Furthermore, it enables configurations with higher $\iota$ to be well replicated with permanent magnets, due to an increase in the available volume for the magnets.

%% file: conclusions.tex
\section{Conclusions}
\label{sec:concl}

Coil optimization, construction, and assembly have proven to be major challenges and sources of cost in the development of modern stellarators \cite{Bosch2013,Lobsien2018,Lion2021,Strykowsky2009}. As interest in stellarators grows, driven by recent promising results, it is important to find budget-efficient ways of building new machines. Considering a large number of tokamaks have been, and will be decommissioned over the years, one attractive possibility would be to convert tokamaks to stellarators. Although research on permanent magnet stellarators has been focused on the development of new machines, due to their versatility, low cost and only requiring TF coils, they would be a promising way of achieving this goal. In this work, the conversion of a specific tokamak, ISTTOK, into a stellarator, is explored.  

In \cref{chapter:designchoices}, several design choices are made using a uniform toroidal grid regarding magnetic configuration, the vessel shape, and the magnet grid. We find the QA configuration to be the most viable one, choosing a circular vessel instead of one that follows the magnetic surface shape. The chosen magnet grid is one with cubic magnets that follow face, face-edge, and face-corner polarizations. In \cref{section:Magpie}, the uniform grid is upgraded to accommodate the required spacing for a mounting structure using cubic magnets with easily achievable orientations, resulting in an increase of one order of magnitude in $f_B$, the used field quality metric.  \cref{section:ISTELL} finds an optimized equilibrium tailored to ISTTOK's characteristics. A QA equilibrium with quasisymmetry residuals of $3.49 \times 10^{-7}$, $\langle \iota \rangle = 0.122$, $A = 10.8$, and that fits ISTTOK's vessel is found. Then, three PM optimization scenarios are evaluated: one maintaining ISTTOK's setup, one removing some coils to increase available volume on the high field side, and one replacing the vacuum vessel. For the latter two scenarios, the plasma accessing ports are modeled, leading to two viable solutions for the conversion of ISTTOK into a stellarator.

While this work has been successful in evaluating the feasibility of converting the ISTTOK tokamak into a stellarator, it is important to recognize that the design and development of a magnetic confinement device is a complex task. Hence, there are multiple remaining steps to be completed.
ISTTOK's diagnostic ports do not follow any symmetries. This means that by rotating the PM grid, the available positions for the magnets change and, therefore so does the optimization result.  Hence, finding the best relative position of the two could be advantageous both for port access and magnet optimization. 
In \cref{section:Magpie} we define the space between magnets and the toroidal spacing by using NCSX as a reference. This risk-averse strategy takes an overestimation of the required spacing in order to guarantee that the magnet grid could be built. However, ISTTOK is a much smaller machine than NCSX. Finding the required spacing to build a support structure, through rigorous structural analysis could considerably improve the error fields obtained in this work.
Finally, it is important to analyze the plasma heating possibilities. ISTTOK currently uses ohmic heating exclusively. A tokamak-stellarator hybrid magnetic equilibrium (see e.g. Ref. \cite{landreman_sengupta_2019}) with an induced plasma current could be considered, keeping the current heating system. This might be an attractive solution in converting a tokamak to a stellarator-like scenario. It would allow higher $\iota$ than what was obtained in the present work while keeping stellarator-like properties. Alternatively, an RF heating system would likely have to be designed and installed.

%% file: acknowledgements.tex
\section*{Acknowledgements}

We thank Alan Kaptanoglu and Ken Hammond for the insightful discussions on GPMO and MAGPIE.
R. J. is supported by the Portuguese FCT-Fundação para a Ciência e Tecnologia, under Grant 2021.02213.CEECIND and DOI  \href{https://doi.org/10.54499/2021.02213.CEECIND/CP1651/CT0004}{10.54499/2021.02213.CEECIND/CP1651/CT0004}.
Simulations were carried out using the EUROfusion Marconi supercomputer facility.
This work has been carried out within the framework of the EUROfusion Consortium, funded by the European Union via the Euratom Research and Training Programme (Grant Agreement No 101052200 - EUROfusion). Views and opinions expressed are however those of the author(s) only and do not necessarily reflect those of the European Union or the European Commission. Neither the European Union nor the European Commission can be held responsible for them.
IPFN activities were supported by FCT - Fundação para a Ciência e Tecnologia, I.P. by project reference UIDB/50010/2020 and DOI  \href{https://doi.org/10.54499/UIDB/50010/2020}{10.54499/UIDB/50010/2020}, by project reference UIDP/50010/2020 and DOI \href{https://doi.org/10.54499/UIDP/50010/2020}{10.54499/UIDP/50010/2020} and by project reference LA/P/0061/202 and  DOI \href{https://doi.org/10.54499/LA/P/0061/2020}{10.54499/LA/P/0061/2020}.